\documentclass[sigconf,authorversion]{acmart}

\includecomment{techreport}
\excludecomment{conference}

\begin{conference}
\setcopyright{acmlicensed}
\end{conference}
\begin{techreport}
\setcopyright{acmlicensed}
\end{techreport}

\copyrightyear{2019}
\acmYear{2019}
\acmConference[AsiaCCS'19]{ACM ASIA Conference on Computer
  and Communications Security}{July 9--12, 2019}{Auckland, New Zealand}
\acmBooktitle{ACM Asia Conference on Computer and Communications Security (AsiaCCS'19), July 9--12, 2019, Auckland, New Zealand}
\acmPrice{15.00}
\acmDOI{10.1145/3321705.3329835}
\acmISBN{978-1-4503-6752-3/19/07}

\begin{conference}
\end{conference}
\begin{techreport}
\end{techreport}

\usepackage{graphicx}

\usepackage{algorithmic}
\usepackage{amsthm}
\usepackage[caption=false,font=footnotesize]{subfig}

\usepackage{dblfloatfix}
\usepackage{url}

\usepackage{comment}
\usepackage{paralist}

\PassOptionsToPackage{hyphens}{url}
\usepackage{hyphenat}

\usepackage{enumitem}

\usepackage{listings}% http://ctan.org/pkg/listings
\usepackage{color}
\usepackage{balance}
\usepackage{tikz}

\DeclareMathAlphabet{\mathcal}{OMS}{cmsy}{m}{n} % replace mathcal default for CCS papers

\newcommand{\circledash}[1]{\tikz \draw[fill=white,white,text=black,inner sep=0,outer sep=0,baseline=O.base,font=\fontsize{8}{\baselineskip}\selectfont] (0,0) circle (0.12cm) node (O) {-};}
\newcommand{\Alice}{\mathrm{Alice}}
\newcommand{\Bob}{\mathrm{Bob}}

\newcommand{\init}{\schemefont{Init}}
\newcommand{\proc}{\schemefont{Proc}}
\renewcommand{\O}{\mathcal{O}}
\newcommand{\st}{\mathit{st}}

\newcommand{\hyb}{\mathrm{hyb}}
\newcommand{\sgx}{\mathrm{sgx}}
\newcommand{\even}{\mathrm{e}}
\newcommand{\odd}{\mathrm{o}}

\newcommand{\inddollar}{\textup{IND\$}}
\newcommand{\sfe}{\textup{SFE}}
\newcommand{\sfep}[1]{\textup{SFE}}
\newcommand{\wsfef}{\textup{SFE\mbox{-}ODD}}
\newcommand{\coins}{w}

\usepackage{amsmath}
\renewcommand{\vec}[1]{\mathbf{#1}}
\newcommand{\dom}{\mathrm{dom}}
\newcommand{\Prob}[1]{{\Pr\hspace{0pt}\left[\,{#1}\,\right]}}
\newcommand{\getsr}{{\:{\leftarrow{\hspace*{-1pt}\raisebox{.75pt}{$\scriptscriptstyle\$$}}}\:}}

\newcommand{\emptystr}{\varepsilon}
\newcommand{\cat}{\,\|\,}
\newcommand{\concat}{\cat}
\newcommand{\bits}{\{0,1\}}

\newcommand{\len}[1]{\schemefont{len}(#1)}

\newcommand{\schemefont}[1]{\scriptsize{\mathsf{#1}}}
\newcommand{\garbler}{\mathcal{G}}
\newcommand{\Gb}{\schemefont{Gb}}
\newcommand{\En}{\schemefont{En}}
\newcommand{\Enl}{\schemefont{EnA}}
\newcommand{\Enr}{\schemefont{EnB}}
\newcommand{\De}{\schemefont{De}}
\newcommand{\Ev}{\schemefont{Ev}}

\newcommand{\tokens}{(X^0_1, X^1_1, \ldots, X^0_{n}, X^1_{n})}

\newcommand{\splitter}{\mathcal{P}}
\newcommand{\idsplitter}{\mathcal{I}}

\newcommand{\splitl}{\schemefont{SpA}}
\newcommand{\splitr}{\schemefont{SpB}}
\newcommand{\gc}{\mathrm{gc}}

\newcommand{\out}{\mathrm{out}}

\newcommand{\exec}{\schemefont{Exec}}

\newcommand{\veca}{\vec{a}}
\newcommand{\vecb}{\vec{b}}
\newcommand{\vecy}{\vec{y}}

\newcommand{\proto}{\mathrm{\Pi}}
\newcommand{\ot}{\mathrm{ot}}
\newcommand{\View}{\schemefont{View}}
\newcommand{\Out}{\schemefont{Out}}

\newcommand{\symenc}{\mathrm{\Gamma}}
\newcommand{\kg}{\mathcal{K}}
\newcommand{\enc}{\mathcal{E}}
\newcommand{\dec}{\mathcal{D}}

\newcommand{\adversaryfont}[1]{\mathcal{#1}}
\newcommand{\advA}{\adversaryfont{A}}
\newcommand{\advB}{\adversaryfont{B}}

\newcommand{\advS}{\adversaryfont{S}}

\newcommand{\cryptofont}[1]{\mathsf{#1}}
\newcommand{\Adv}[1]{\cryptofont{Adv}^{\mathrm{\scriptsize \MakeLowercase{#1}}}}
\newcommand{\Exp}[1]{\cryptofont{Exp}^{\mathrm{\scriptsize \MakeLowercase{#1}}}}

\newcommand{\gamesfontsize}{\small}
\newcommand{\outputs}{=}
\newcommand{\true}{1}
\newcommand{\false}{0}
\newcommand{\Foreach}[3]{for $#1 \gets #2$ to $#3$ do}
\newcommand{\ind}{\hspace*{10pt}}
\newcommand{\pick}{\schemefont{pick}}
\newcommand{\guess}{\schemefont{guess}}
\definecolor{CommentColor}{RGB}{125,175,230}
\newcommand{\codecomment}[1]{\textcolor{CommentColor}{\,\textbf{\it\#}\,#1}}

\newcommand{\oneCol}[2]{
\begin{center}
        \framebox{
        \begin{tabular}{c@{\hspace*{.4em}}}
        \begin{minipage}[t]{#1\textwidth}\gamesfontsize
          #2
        \end{minipage}
        \end{tabular}
        }
\end{center}
}

\newcommand{\twoColsTwoRows}[5]{
  \providecommand{\pad}{6pt}
  \makebox[\textwidth][c]{
  \begin{tabular}{|@{\hskip \pad}l@{}|@{}@{\hskip \pad}l|}
    \hline
    \rule{0pt}{1\normalbaselineskip}
    \begin{minipage}[t]{#1\textwidth}\gamesfontsize
      #2 \vspace{\pad}
    \end{minipage} &
    \begin{minipage}[t]{#1\textwidth}\gamesfontsize
      #3 \vspace{\pad}
    \end{minipage} \\
    \hline
    \rule{0pt}{1\normalbaselineskip}
    \begin{minipage}[t]{#1\textwidth}\gamesfontsize
      #4 \vspace{\pad}
    \end{minipage} &
    \begin{minipage}[t]{#1\textwidth}\gamesfontsize
      #5 \vspace{\pad}
    \end{minipage} \\
    \hline
  \end{tabular}
  }
}

\renewcommand{\paragraph}[1]{\vspace{0.5em}\noindent{\sc #1. }}

\definecolor{darkgreen}{RGB}{50,127,0}
\definecolor{extracolor}{rgb}{1,.3,0}

\newcommand{\tom}[1]{{\color{magenta} \em(#1 --\textit{Tom})}}

\newcommand{\xfont}[1]{\ifmmode\mathrm{#1}\else{#1}\fi}
\newcommand{\oldx}{\xfont{na{\"i}ve SGX-enabled SFE}}
\newcommand{\Oldx}{\xfont{Na{\"i}ve SGX-enabled SFE}}
\newcommand{\newx}{\xfont{SGX-enabled SFE}}
\newcommand{\Newx}{\xfont{SGX-enabled SFE}}
\newcommand{\mixx}{\xfont{hybrid SFE-SGX}}
\newcommand{\Mixx}{\xfont{Hybrid SFE-SGX}}
\newcommand{\gcx}{\xfont{GC-based SFE}}
\newcommand{\Gcx}{\xfont{GC-based SFE}}

\begin{document}

\title{A Hybrid Approach to\\Secure Function Evaluation using SGX} % TODO: replace with your title
%\begin{techreport}
%\titlenote{Full version, with proofs, of conference paper at AsiaCCS 2019~\cite{this_conf}.}
%\end{techreport}

% author block would go here
%
% The "author" command and its associated commands are used to define the authors and their affiliations.
% Of note is the shared affiliation of the first two authors, and the "authornote" and "authornotemark" commands
% used to denote shared contribution to the research.

\author{Joseph I. Choi}
\email{choijoseph007@ufl.edu}
\affiliation{University of Florida}

\author{Dave (Jing) Tian}
\email{daveti@ufl.edu}
\affiliation{University of Florida}

\author{Grant Hernandez}
\email{grant.hernandez@ufl.edu}
\affiliation{University of Florida}

\author{Christopher Patton}
\email{cjpatton@ufl.edu}
\affiliation{University of Florida}

\author{Benjamin Mood}
\email{bmood@pointloma.edu}
\affiliation{Point Loma Nazarene University}

\author{Thomas Shrimpton}
\email{teshrim@ufl.edu}
\affiliation{University of Florida}

\author{Kevin R. B. Butler}
\email{butler@ufl.edu}
\affiliation{University of Florida}

\author{Patrick Traynor}
\email{traynor@ufl.edu}
\affiliation{University of Florida}

% By default, the full list of authors will be used in the page headers. Often, this list is too long, and will overlap
% other information printed in the page headers. This command allows the author to define a more concise list
% of authors' names for this purpose.
\renewcommand{\shortauthors}{J. I. Choi et al.}

\begin{abstract}

  A protocol for two-party secure function evaluation (2P-SFE) aims to allow the
  parties to learn the output of function~$f$ of their private inputs, while
  leaking nothing more.
  In a sense, such a protocol realizes a trusted oracle that computes~$f$ and
  returns the result to both parties.
  There have been tremendous strides in efficiency over the past ten years, yet
  2P-SFE protocols remain impractical for most real-time, online computations,
  particularly on modestly provisioned devices.
  Intel's Software Guard Extensions (SGX) provides hardware-protected
  execution environments, called \emph{enclaves}, that may be viewed
  as trusted computation oracles.
  %
  %However, this trust is somewhat misplaced, since these devices expose side-channels in practice.
  While SGX provides native CPU speed for secure computation,
  previous side-channel and micro-architecture attacks have demonstrated
  how security guarantees of enclaves can be compromised.
  
  In this paper,
  we explore a balanced approach to 2P-SFE on SGX-enabled processors
  by constructing
  a protocol for evaluating~$f$ \emph{relative to a partitioning of~$f$.}
  %of the
  %function (and the parties' inputs) into multiple rounds, some to be evaluated
  %within an SGX enclave, and the others via traditional cryptographic methods.
  %
  This approach alleviates the burden of trust on the enclave by allowing the
  protocol designer to choose which components should be evaluated within the
  enclave, and which via standard cryptographic techniques.
  We describe SGX-enabled SFE protocols (modeling the enclave as an oracle),
  and formalize the strongest-possible notion of 2P-SFE for our setting. We
  prove our protocol meets this notion when properly realized.
  \if{0}
  ---~in general, traditional 2P-SFE is not achievable when the
  computation of~$f$ is partitioned and intermediate function values are
  observed.  We prove that our protocol meets this notion, when properly
  realized.
  \fi
  We implement the protocol and apply it to two practical problems:
  privacy-preserving queries to a database, and a version of Dijkstra's
  algorithm for privacy-preserving navigation.  Our evaluation shows
  that our SGX-enabled SFE scheme enjoys a 38x increase in performance over
  garbled-circuit-based SFE.
  Finally, we justify modeling of the enclave as an oracle by implementing
  protections against known side-channels.
\end{abstract}

\if{0}
Unfortunately, na\"ively implementing such protocols on
SGX-enabled devices creates vulnerabilities.\tom{like what?  Can we give some suggestive words here?}

In this paper, we provide a provable-security treatment for SGX-enabled 2P-SFE.  We formalize two-party protocols with an explicit oracle, which may share a secret key with one of the parties.  Intuitively, this oracle abstracts the SGX-enclave.  We also formalize what it means to partition a function~$f$ (the target of the 2P-SFE) into subfunctions, capturing the idea of computing some portion of~$f$ within the SGX-enclave, and some portion outside of it.  We prove that a partitioning scheme for~$f$, a symmetric encryption scheme, a garbling scheme and a two-party oblivious transfer scheme can be composed in a straightforward manner to achieve secure SGX-enabled 2P-SFE.  Supporting our formal modeling of the SGX-enclave as an oracle, we consider (and instantiate) methods for closing known side-channels that would invalidate the black-box behavior of an oracle.

Our implementation of a hybrid 2P-SFE scheme on SGX-enabled hardware
demonstrates an increase in performance of up to 38X over a garbled
circuit-based approach while providing provable security guarantees about
the scheme and defending against control-channel attacks that target SGX.
These improvements provide near real-time performance for 2P-SFE,
dramatically increasing its practicality for use in systems.
\fi




%\ccsdesc{Security and privacy~Use https://dl.acm.org/ccs.cfm to generate actual concepts section for your paper}
% -- end of section to replace with generated code

%
% Keywords. The author(s) should pick words that accurately describe the work being
% presented. Separate the keywords with commas.

% A "teaser" image appears between the author and affiliation information and the body 
% of the document, and typically spans the page. 
%  \Description{Enjoying the baseball game from the third-base seats. Ichiro Suzuki preparing to bat.}

\maketitle

\section{Introduction}
Secure function evaluation (SFE) describes the process of multiple parties
collectively computing a function and receiving its output without learning the
inputs from any other party. Originally proposed in the 1980s, SFE was primarily
of theoretical interest until the 2000s, when practical implementations of
two-party SFE (2P-SFE) became available. Since then, interest in the space has
dramatically increased and the costs of computation have been lowered by orders
of magnitude.

Despite this success in reducing the costs of SFE, it is still not yet
sufficiently practical to be used in applications where (near) real-time
performance is required. This is in large part due to the substantial number of
cryptographic operations that the parties need to perform.  In the 2P-SFE case,
this is often manifested by representing the function to be computed over as a
circuit and {\em garbling} all of its input and output wires, as well as truth
tables associated with each logic gate.

Hardware support for secure computing offers a chance to reduce these costs.
Specifically, Intel's Software Guard Extensions (SGX) provides secure memory regions
(called enclaves) inside which code and data can live outside of the purview of the
operating system or system administrator.  This platform thus offers the
potential to enable SFE without the often crippling overheads of the associated
cryptographic constructions.
%However, controlled-channel
%attacks~\cite{xu2015controlled} and recent work demonstracing
%exfiltration of private keys through side channels~\cite{malwaregx}
%demonstrate that the na\"{i}ve use of such hardware can lead to the exposure of
%sensitive information. Intel themselves explicitly place the onus of
%preventing side channel attacks on the enclave
%developer~\cite{stancesc}, yet side channels are left out of scope by
%previous work attempting to model SGX operations~\cite{sinha2015moat}.
%% High-level computational model.
%%\paragraph{Our contributions}
%%%% TODO CITE OTHER: fix this paragraph
%The constraints of enclave memory considerations have similarly been ignored
%by previous work aimed at coupling SGX
%processor extensions with SFE~\cite{smcsgx-other}.  Recent work tends to na\"{i}vely place entire
%programs into an enclave, which is then trusted without question.
While SGX provides native CPU speed for secure computation,
previous side-channel and micro-architecture attacks have demonstrated how
security guarantees of enclaves can be compromised.
Controlled-channel attacks~\cite{xu2015controlled} leverage the page fault handler to leak sensitive information inside an enclave.
Leaky Cauldron~\cite{leaky} shows memory side channel hazards in SGX ranging from TLB to DRAM modules.
Meltdown~\cite{meltdown} and Spectre~\cite{spectre} attacks can also be applied to enclaves,
and Foreshadow~\cite{foreshadow} has successfully extracted the CPU attestation key from enclaves thus breaking the SGX remote attestation.
Some of these attacks could be mitigated by microcode update or system hardening.
However, they do expose two important questions:

\begin{enumerate}
\item \textit{Is it reasonable to put all secrets into an enclave?}
\item \textit{What could we do if SGX might be compromised?}
\end{enumerate}

This paper goes beyond simply introducing SGX to SFE~\cite{smcsgx-other}.
Rather, in addition to reducing the computation required for 2P-SFE operations,
our scheme provides provable assurance that the most sensitive inputs
of either party will remain protected.
%This paper represents a new approach to 2P-SFE.
%by introducing the use of SGX
%processor extensions to reduce the computation required for these operations.
In summary, our main contributions\footnote{Our source code will be made
available on~\url{https://github.com/FICS/smcsgx}.} are as follows:

\begin{itemize}
  \item \textbf{Design of Hybrid SFE Scheme:} We provide a new construction that
    considers the partitioning of a function into multiple segments, some to be
    executed within an SGX enclave and the others within a garbled circuit.  We
    surface the idea of partitioning a function~$f$ as a first-class primitive
    in the design, as different partitionings induce different schemes, with
    different efficiency and trust assumptions.

  \item \textbf{Formal Protocol Analysis:} We formalize two-party protocols with
    an explicit oracle, to support analysis of SFE protocols that leverage an
    SGX enclave.  We also formalize a best-possible notion of 2P-SFE for our
    setting~--- traditional 2P-SFE is not possible, in general~--- and prove
    that our Hybrid SFE scheme, properly realized, meets this notion.
%
%  \item \textbf{Consideration of Side-Channels:} We introduce a scheme that
%    executes solely within the enclave but with side channel protections.  These
%    protections also support our formal modeling of the SGX-enclave as a
%    black-box oracle.

  \item \textbf{Evaluation and Case Studies:} We provide practical scenarios for
    using our SFE schemes, including a database interacting with a client for
    privacy-preserving queries and a location-finding scenario with a
    privacy-preserving Dijkstra's algorithm. Our extensive evaluation
    demonstrates the overhead of these operations and shows that in the hybrid
    SFE scheme, we increase performance by up to 38x compared to garbled
    circuits. We also provide an empirical demonstration of resilience to
    controlled-channel attacks.
\end{itemize}

%
% daveti: start-moving-to-new-background
%
\if{0}
SGX provides a very efficient platform for secure computation. However,
the trust model associated with SGX is substantially different than with
garbled circuits or other cryptographic approaches to SFE. The secure
enclave in which the computation is being performed must attest to
the remote party providing data that it is trustworthy. To do this, the
enclave must be issued an identifier by Intel and must use an attestation
service provisioned by Intel. While these services have been described at a
high level~\cite{intelepid}, they have not been formally investigated or
subjected to intensive external review by the security community.
Furthermore, the size of the enclave is limited by the enclave page cache, which
is limited to approximately 128 MB. 
\footnote{128 MB may be sufficient to run a single application that is not memory-intensive,
but it cannot support entire web servers or databases with substantial memory usage.
It is especially not enough for a cloud environment, in which multiple subscribers
must share the EPC memory of a cloud server.
Running multiple full applications would lead to expensive process-swapping operations,
negatively impacting performance. Besides these performance considerations, the Intel SGX SDK 
requires developers to partition programs in order to reduce the
TCB contribution of the enclave.}
For these reasons, parties may not want to
rely solely on SGX to compute a function~$f$ on their private inputs.  However,
traditional approaches to SFE using garbled circuits, even using the most
efficient schemes, are still orders of magnitude slower than performing the same
operations non-securely. %%% TODO CITE OTHER: naive paper how they suck, partitioning
\footnote{In certain scenarios, secure computation based
on the GMW construction may outperform garbled circuit
execution~\cite{Schneider2013-ma} but the larger point still remains about the
substantial overhead incurred by privacy-preserving schemes.}

\if{0}
  Bahmani et. al~\cite{smcsgx-other} previously attempted to achieve more
  efficient secure computation by solely relying on SGX. However, their approach
  is marked by a number of shortcomings. They propose placing entire programs
  into an SGX enclave, a practice that lies in direct conflict with Intel's
  programming paradigm.  Doing so does not make code any more secure, nor is it
  practical, given the limitations of enclave page cache memory.  Furthermore,
  their protocol requires involved parties to na\"{i}vely place complete trust
  in the enclave.
\fi

We consider the middle ground: parts of the computation of~$f$ are performed using
SGX, and parts are performed using traditional SFE mechanisms (e.g., garbled
circuits and oblivious transfer). Exactly \emph{how} one partitions the function
is an efficiency- and security-critical matter.  The efficiency viewpoint has
already been touched upon, so we take up the security viewpoint.

%Specifically, we consider partitioning a function~$f$ into a sequence of
%functions~$f_1,f_2,\ldots,f_\ell$ (with corresponding partitioning of the
%parties' private inputs), where the odd-indexed functions are computed
%by~$\Alice$, and the even-indexed functions by~$\Bob$.  The partitioning is
%asymmetric, in the sense that odd-indexed functions take three inputs ---~one
%from~$\Alice$, one from~$Bob$, and some state~--- and return two values, while
%even-indexed functions take only inputs from~$\Alice$ and~$\Bob$, and return a
%single value.  Intuitively, the odd-indexed functions are capturing
%computations within an SGX enclave, which may carry state across its
%executions, and even-indexed functions capture non-enclave (and stateless)
%computations.

\begin{figure}[t]
\centering
  \includegraphics[scale=0.82]{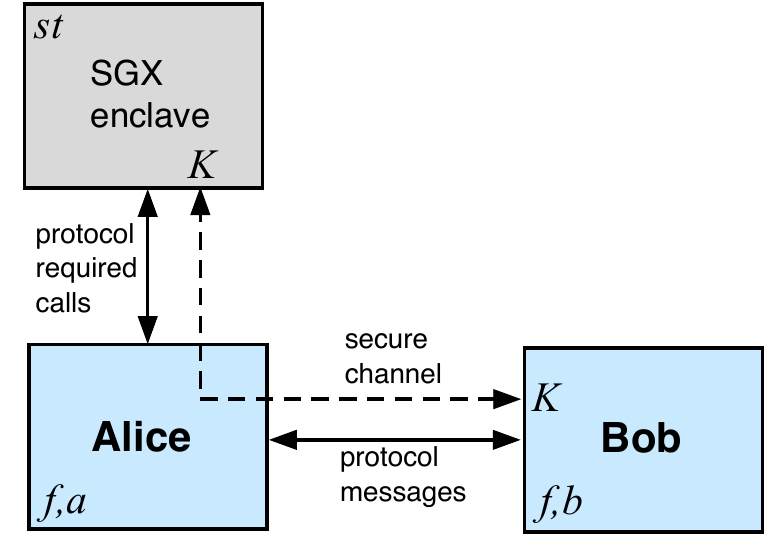}
\caption{High-level view of our execution model.}
%\includegraphics[scale=.4]{figures/hybrid_intro.pdf}
%\caption{High-level picture of our \Mixx\ protocol}
\label{fig:hybrid-intro}
\end{figure}

%\paragraph{Security implications of partitioning}
Our setting is loosely captured in Figure~\ref{fig:hybrid-intro}.  Alice and Bob
would like to carry out a protocol for 2P-SFE of $f(a,b)$.  Alice has black-box
access to an SGX enclave that her environment hosts. (We will justify this
black-box modeling in a moment.) Bob, the remote party, shares a secret key with
the enclave, so that he may communicate with the enclave via a secure channel;
Alice has view of this channel, but does not possess the key.  As Alice and Bob
carry out the protocol, Alice will make ``queries'' to her enclave, asking it to
compute intermediate functions that are specified by a given partitioning
of~$f$.  She provides private input (related to~$a$) as part of these queries,
and Bob provides private input (related to~$b$) via the secure channel.  The
results of these enclave queries are visible to Alice, and used in subsequent
parts of the protocol for computing~$f(a,b)$.

Thus, intermediate information about the computation of~$f$ is leaked to Alice,
even when she participates honestly. This implies that, in general, the
standard notion of 2P-SFE \emph{is not possible} in this setting.  A bad
partitioning of~$f$ may result in intermediate values that leak information
about private inputs~$a$ and~$b$ of Alice and Bob.
%%% TODO CITE OTHER: can we point to other paper show that they don't do hybrid

%The standard SFE notion formalizes a party's ability to discern
%information about the other player's input beyond that which follows from the
%result. This notion is out of reach in our setting, since the output of the SGX
%potentially provides Alice with additional information about Bob's input.

We define, and aim to achieve, the best possible SFE notion in this setting.
Namely, to show that a protocol for computing~$f(a,b)$ leaks nothing more than
$f(a,b)$ and the intermediate values.  That is, 2P-SFE with respect to a
\emph{particular way of partitioning the function.}

We formally model an SGX enclave as a black-box, but in reality an enclave may
leak information about the computations they perform due to timing side channels,
memory access patterns, and controlled-channel attacks on program
flow~\cite{xu2015controlled}.
Enclave malware was recently demonstrated to be capable of cache attacks
on co-located enclaves, recovering nearly entire RSA private keys within
5 minutes~\cite{malwaregx}. To support our black-box modeling of an enclave, we 
include mitigations for side-channels in our implementations.
%our implementations contain mitigations for
%these.

\fi
%
% daveti: end-moving-to-new-background
%

%% TODO: THIS IS MESSED UP FIX
\paragraph{Outline} The rest of the paper proceeds as follows:
%
%Section~\ref{sec:sgx-background} provides background on SGX;
Section~\ref{sec:back2} provides a high-level view of our hybrid approach;
Section~\ref{sec:prelims} provides preliminary notation;
Section~\ref{sec:partitioning} provides a detailed formal treatment of
SGX-enabled SFE;
Section~\ref{sec:hybrid} defines and analyzes our hybrid protocol;
Section~\ref{sec:side-channel} discusses side-channels and mitigations for our
implementation;
Section~\ref{sec:design} presents the design and implementation of the protocol
schemes within a real SGX environment;
Section~\ref{sec:eval} describes our experimental evaluation;
Section~\ref{sec:relwork} considers related work; and
Section~\ref{sec:conc} concludes.

\section{Hybrid Approach High-level View}
\label{sec:back2}

SGX provides a very efficient platform for secure computation. However,
the trust model associated with SGX is substantially different than with
garbled circuits or other cryptographic approaches to SFE. The secure
enclave in which the computation is being performed must attest to
the remote party providing data that it is trustworthy. To do this, the
enclave must be issued an identifier by Intel and must use an attestation
service provisioned by Intel.
%While these services have been described at a
%high level~\cite{intelepid}, they have not been formally investigated or
%subjected to intensive external review by the security community.
Unfortunately, Foreshadow attacks have demonstrated that
it is hard to guarantee that the sensitive code and data are running in a real enclave
rather than an emulation environment even if SGX remote attestation succeeds,
leaving alone threats from other side-channel and micro-architecture attacks.
Furthermore, the size of the enclave is limited by the enclave page cache, which
is limited to approximately 128 MB.% 
\footnote{128 MB may be sufficient to run a single application that is not memory-intensive,
but it cannot support entire web servers or databases with substantial memory usage.
It is especially not enough for a cloud environment, in which multiple subscribers
must share the EPC memory of a cloud server.
Running multiple full applications would lead to expensive process-swapping operations,
negatively impacting performance. Besides these performance considerations, the Intel SGX SDK 
requires developers to partition programs in order to reduce the
TCB contribution of the enclave.}
For these reasons, parties may not want to
rely solely on SGX to compute a function~$f$ on their private inputs.  However,
traditional approaches to SFE using garbled circuits, even using the most
efficient schemes, are still orders of magnitude slower than performing the same
operations non-securely.% %%% TODO CITE OTHER: naive paper how they suck, partitioning
\footnote{In certain scenarios, secure computation based
on the GMW construction may outperform garbled circuit
execution~\cite{Schneider2013-ma} but the larger point still remains about the
substantial overhead incurred by privacy-preserving schemes.}

\if{0}
  Bahmani et. al~\cite{smcsgx-other} previously attempted to achieve more
  efficient secure computation by solely relying on SGX. However, their approach
  is marked by a number of shortcomings. They propose placing entire programs
  into an SGX enclave, a practice that lies in direct conflict with Intel's
  programming paradigm.  Doing so does not make code any more secure, nor is it
  practical, given the limitations of enclave page cache memory.  Furthermore,
  their protocol requires involved parties to na\"{i}vely place complete trust
  in the enclave.
\fi

We consider the middle ground: parts of the computation of~$f$ are performed using
SGX, and parts are performed using traditional SFE mechanisms.
%(e.g., garbled
%circuits and oblivious transfer). 
Exactly \emph{how} one partitions the function
is an efficiency- and security-critical matter.  The efficiency viewpoint has
already been touched upon, so we take up the security viewpoint.

%Specifically, we consider partitioning a function~$f$ into a sequence of
%functions~$f_1,f_2,\ldots,f_\ell$ (with corresponding partitioning of the
%parties' private inputs), where the odd-indexed functions are computed
%by~$\Alice$, and the even-indexed functions by~$\Bob$.  The partitioning is
%asymmetric, in the sense that odd-indexed functions take three inputs ---~one
%from~$\Alice$, one from~$Bob$, and some state~--- and return two values, while
%even-indexed functions take only inputs from~$\Alice$ and~$\Bob$, and return a
%single value.  Intuitively, the odd-indexed functions are capturing
%computations within an SGX enclave, which may carry state across its
%executions, and even-indexed functions capture non-enclave (and stateless)
%computations.

\begin{figure}[t]
\centering
  \includegraphics[scale=0.82]{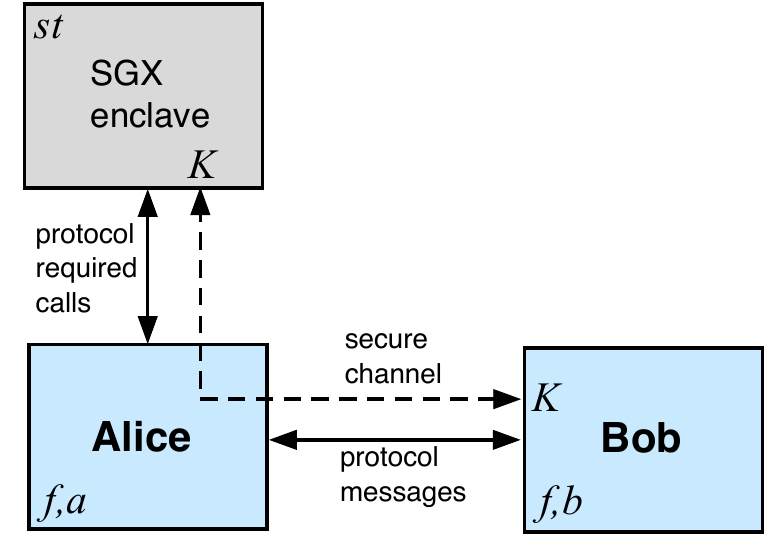}
\caption{High-level view of our execution model.}
%\includegraphics[scale=.4]{figures/hybrid_intro.pdf}
%\caption{High-level picture of our \Mixx\ protocol}
\label{fig:hybrid-intro}
\end{figure}

%\paragraph{Security implications of partitioning}
Our setting is loosely captured in Figure~\ref{fig:hybrid-intro}.  Alice and Bob
would like to carry out a protocol for 2P-SFE of $f(a,b)$.  Alice has black-box
access to an SGX enclave that her environment hosts. (We will justify this
black-box modeling in a moment.) Bob, the remote party, shares a secret key with
the enclave,
used to establish a secure communication channel; 
%for communication via a secure channel;
%so that he may communicate with the enclave via a secure channel;
Alice has view of this channel, but does not possess the key.  As Alice and Bob
carry out the protocol, Alice will make ``queries'' to her enclave, asking it to
compute intermediate functions that are specified by a given partitioning
of~$f$.  She provides private input (related to~$a$) as part of these queries,
and Bob provides private input (related to~$b$) via the secure channel.  The
results of these enclave queries are visible to Alice, and used in subsequent
parts of the protocol for computing~$f(a,b)$.

Thus, intermediate information about the computation of~$f$ is leaked to Alice,
even when she participates honestly. This implies that, in general, the
standard notion of 2P-SFE \emph{is not possible} in this setting.  A bad
partitioning of~$f$ may result in intermediate values that leak information
about private inputs~$a$ and~$b$ of Alice and Bob.
%%% TODO CITE OTHER: can we point to other paper show that they don't do hybrid

%The standard SFE notion formalizes a party's ability to discern
%information about the other player's input beyond that which follows from the
%result. This notion is out of reach in our setting, since the output of the SGX
%potentially provides Alice with additional information about Bob's input.

We define, and aim to achieve, the best possible SFE notion in this setting.
Namely, to show that a protocol for computing~$f(a,b)$ leaks nothing more than
$f(a,b)$ and the intermediate values.  That is, 2P-SFE with respect to a
\emph{particular way of partitioning the function.}

We formally model an SGX enclave as a black-box, but in reality, an enclave may
leak information about the computations it performs due to timing side channels,
memory access patterns, and controlled-channel attacks on program
flow~\cite{xu2015controlled}.
Enclave malware was recently demonstrated to be capable of cache attacks
on co-located enclaves, recovering nearly entire RSA private keys within
5 minutes~\cite{malwaregx}. To support our black-box modeling of an enclave, we 
include mitigations for side-channels in our implementations.
%our implementations contain mitigations for
%these.

\newcommand{\event}{\mathsf{E}}
\section{Preliminaries}
\label{sec:prelims}
In order to design a secure hybrid scheme using SGX,
we first introduce the fundamental protocols and their associated notations.
This includes our main protocol involving two parties Alice and Bob,
an introduction to Garbled Circuit (GC) syntax, and finally \emph{1-2 oblivious
transfer} in the context of this work.\\[0.2cm]
%
%The inputs and outputs of a function are fixed-length strings (or tuples over
%fixed-length strings) unless otherwise stated.
%
Let $\dom f$ denote the domain of a function $f$.
We use~$\emptystr$ to denote the empty string. If $a$ and $b$ are strings, let
$a \cat b$ denote their concatenation.
%
%If $\veca$ is a vector, let $|\veca|$ denote its length and let $\veca[i]$
%denote its $i$-th element.
%
Let $y \getsr A(x_1, \ldots)$ denote the execution of a randomized algorithm~$A$
on input~$(x_1, \ldots)$ and assigning of the output to~$y$. We write
$y \gets A(x_1, \ldots)$ if~$A$ is deterministic.
Algorithms are randomized unless noted otherwise.
%
%We use the notation $\Prob{\op_1; \op_2; \cdots; \op_t \,:\, \event}$ to denote
%the probability of event~$\event$ occurring as a result of executing
%operations $\op_1,\op_2,\ldots,\op_t$ in sequence.

%%%%%%%%%%%%%%%%%%%%%%%%%%%%%%%%%%%%%%%%%%%%%%%%%%%%%%%%%%%%%%%%%%%%%%%%%%%%%%%
\subsection{Protocols}
An \emph{oracle-relative, two-party protocol}~$\proto$ is a two-party protocol
played by~$\Alice$ and~$\Bob$, in which one or both may have access to
an explicitly defined oracle~$\O$.
The parties and the oracle all have local, private state, which includes a
\emph{long-term input} that is determined during protocol initialization and
accessible across protocol executions.
In our SGX-enabled protocols, the oracle abstracts an SGX enclave that is part
of $\Alice$'s environment, and the long-term inputs of $\Bob$ and the oracle
encode a shared key.
\if{0}
  The latter abstracts the outcome of the SGX enclave and $\Bob$ having executed
  a remote attestation procedure.
\fi

Protocols are executed with respect to players' private inputs. Executing the
protocol on $(a,b)$ means to initiate $\Alice$'s state with private
input~$a$ and $\Bob$'s state with private input~$b$, and exchange messages
until both players halt. (We assume each player and the oracle have been
provisioned with their long-term inputs.)
This is denoted
$
  (y_0, y_1, \pi, \st^\prime) \getsr \proto(1^k, a, b, \st)
$
where~$\st$ is the initial state of~$\O$,
$y_0$ and~$y_1$ are the final states of~$\Alice$ and~$\Bob$ respectively, and
$\st^\prime$ is the final state of~$\O$.
% Transcript
String~$\pi$ is a ``transcript'' of the protocol execution.\footnote{%
The transcript serves no functional purpose, so we omit the details here.}
%
%(The transcript serves no functional purpose, so we omit the details here.)
%
$\Out_{\proto,k}^i(a, b, \st)$ is the random variable denoting the
final state of player~$i$ when executing protocol~$\proto$ on 
%private inputs
$(a,b)$ with $\st$ as the oracle's initial state. (Integer~$k$ is the
security parameter.)
%\begin{conference}
%Refer to Appendix A.1 of our full paper~\cite{this_full}
%for more details.
%\end{conference}
%\begin{techreport}
Refer to Appendix~\ref{sec:proto-syntax} for more details.
%\end{techreport}

%\paragraph{The power of an oracle}
%Even before giving a definition of secure function evaluation in our setting,
%readers familiar with the standard notion may observe the following:
Note that
our protocol syntax
% for providing access to an oracle
admits trivially secure SFE
protocols.  For us, the oracle provides a way to abstractly capture an SGX
enclave, and our formalization provides a convenient way to reason about
SGX-enabled SFE protocols.\footnote{%
Although we choose to reason about SGX-enabled SFE protocols with access to an oracle,
it is also possible to prove composition for protocols without oracles.}
\if{0}
  In addition, the long-term inputs abstractly capture keys shared between the
  enclave and party that have carried out by the SGX remote-attestation protocol
  with it.
  For example, when $\Alice$'s long-term input is $\emptystr$ and $\Bob$ and the
  oracle share a long-term input~$K$, then $\Bob$ may communicate with the
  oracle (i.e., the enclave) via a secure channel leveraging the shared key~$K$;
  whereas $\Alice$ may not.
\fi

\paragraph{Adversarial model}
We will consider the security of oracle protocols in the presence of
\emph{semi-honest} (sometimes called \emph{honest-but-curious}) adversaries.
This means each player executes the protocol faithfully, but may otherwise act
arbitrarily to violate security.  All of our notions ask the adversary to
distinguish its view of the protocol from the output of a simulator, which is
given the private input of the corresponding player (and only the length of the
other player's private input).
Although a \emph{semi-honest} model may not be sufficient for certain classes of 
real-world applications, it offers a natural first step towards SGX-enabled SFE.
In many situations, we can expect computing parties to have a mutual interest
in fulfilling the protocol correctly. 
Our model nevertheless does not prevent
an adversary from attempting to break into the enclave.

%%%%%%%%%%%%%%%%%%%%%%%%%%%%%%%%%%%%%%%%%%%%%%%%%%%%%%%%%%%%%%%%%%%%%%%%%%%%%%%
\subsection{Garbling schemes}
We adopt the syntax of Bellare, Hoang, and Rogaway~\cite{garbling} for
garbled circuits.
\if{0}
  \footnote{To avoid developing additional notation, all garbling scheme
  algorithm inputs and outputs are implicitly encoded as strings. Likewise, we
  silently treat functions $f$, and evaluations of $f$ on inputs, as strings
  (relative to the same implicit alphabet) where syntactically necessary.}
\fi
A \emph{garbling scheme} is a quadruple of algorithms $\garbler = (\Gb, \En, \De,
\Ev)$; the first is randomized, while the rest are deterministic.
The \emph{garbling algorithm}~$\Gb$ takes as input~$1^k$ and a function~$f$, and
outputs a triple of strings $(F, e, d)$. This is written $(F, e, d) \getsr
\Gb(1^k, f)$.
String~$e$ describes the \emph{encoding function}, and
$X \gets \En(e, x)$ denotes the encoding of~$x$ under~$e$; we call~$X$
the \emph{garbled input}.
String~$F$ describes the \emph{garbled function}, and $Y \gets \Ev(F,
X)$ denotes evaluating~$F$ on~$X$, yielding the
\emph{garbled output}~$Y$.
Finally, string~$d$ describes the \emph{decoding function}, and $y
\gets\De(d, Y)$ denotes the decoding of~$Y$ under~$d$, yielding the
\emph{final output}~$y$.
% Correctness
The garbling scheme $\garbler$ is \emph{correct} if for every function $f$,
for every $x$ in the domain of $f$, and for every $(F, e, d)$ in the range of
$\Gb(1^k, f)$, it holds that $f(x) = \De(d, \Ev(F, \En(e, x)))$.

% Projective garbling
Garbling circuits were introduced by Andrew Yao~\cite{Yao1982}. His construction,
as well as most recent designs, have an additional syntactic property necessary for
SFE.
A garbling scheme is called \emph{projective} if the encoding algorithm may be
written as a pair of algorithms $(\Enl, \Enr)$ such that $\En(e, (a,b)) =
\Enl(e, a) \cat \Enr(e, b)$ for every $(a,b) \in \dom f$.
%
%\begin{conference}
%(See Appendix A.2 of our full paper~\cite{this_full} for a more formal definition.)
%\end{conference}
%\begin{techreport}
(See Appendix~\ref{sec:garbler} for a more formal definition.)
%\end{techreport}

Bellare, Hoang, and Rogaway~\cite{garbling} formalize a security notion for
garbling schemes, which we will use here.
The \emph{privacy} of a garbling scheme captures the adversary's ability to
discern anything about~$f$ or~$x$ given only~$F$,~$X$, and~$d$.
The notion is parameterized by \emph{side information} about the function~$f$ ---
for example, the length of its encoding, or the topology of the circuit used to
compute it. The adversary is given the side information as input. In the SFE
setting, the side information is~$f$ itself.
%
%\begin{conference}
%We formalize the simulation-based privacy notion of~\cite{garbling} in
%Appendix A.2 of our full paper~\cite{this_full}.
%\end{conference}
%\begin{techreport}
We formalize the simulation-based privacy notion of~\cite{garbling} in
Appendix~\ref{sec:garbler}.
%\end{techreport}

%%%%%%%%%%%%%%%%%%%%%%%%%%%%%%%%%%%%%%%%%%%%%%%%%%%%%%%%%%%%%%%%%%%%%%%%%%%%%%%
\subsection{1-2 oblivious transfer}
A standard way of facilitating secure multiparty computation (and achieving SFE in
particular) is to compose a projective garbling scheme with an oblivious transfer
protocol  \cite{rabin2005exchange}.
A 1-2 (one-of-two) transfer protocol is a two-party protocol in which~$\Alice$
possesses two equal-length strings~$X^0$ and~$X^1$ and~$\Bob$ possesses a
bit~$b$.
Roughly speaking, a 1-2 transfer protocol is \emph{oblivious} if~$\Bob$
learns~$X^b$ (and only $X^b$) and $\Alice$ learns nothing.
More generally, $\Alice$ may possess a sequence of strings $\tokens$ and $\Bob$
a string~$b$ of length~$n$, and the goal is to obliviously transfer
$(X_1^{b_1},\ldots, X_n^{b_n})$ to~$\Bob$.
%
%\begin{conference}
%We formalize this security property in Appendix A.3 of our full paper~\cite{this_full}.
%\end{conference}
%\begin{techreport}
We formalize this security property in Appendix~\ref{sec:ot}.
%\end{techreport}

\section{SFE for Partitioned Functions}\label{sec:partitioning}
We next describe our construction of SFE for partitioned functions.

Our main protocol partitions the computation of a function into a sequence of
round functions, each depending on a piece of the players' private input.
Odd-round functions are evaluated within $\Alice$'s enclave and may depend on
%the enclave's 
its state.
Even-round functions are stateless and evaluated using standard
cryptographic techniques.

We first give the syntax of an partitioning scheme that captures this
computational model.\footnote{Our aim is to demonstrate the possibility of combining SGX
and Garbled Circuits for SFE, given a proper partitioning scheme.
%We do not discount the difficulty or importance of program partitioning.
Partitioning is itself a hard problem that we do not attempt to solve;
entire papers~\cite{zdancewic, partition_web} have been dedicated to it.
Four possible partitioning schemes for SGX programs are explored
by Atamli-Reineh and Martin~\cite{partition_sgx_casestudy}, ranging from
the na\"ive placement of entire applications in an enclave to separating out
sensitive components into individual enclaves.}
%Other works tackling
%secure program partitioning include~\cite{zdancewic, partition_web}.}
%
We then define secure evaluation of a function relative to a partitioning of it,
which captures the intermediate results available to $\Alice$ and $\Bob$.

\begin{figure}[t]
\centering
\includegraphics[scale=0.82]{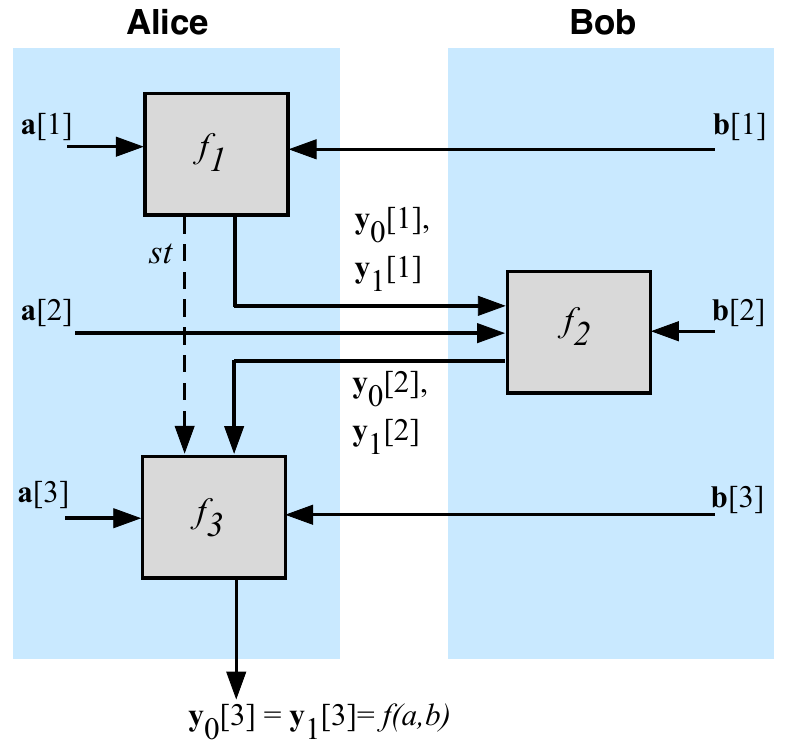}
  \caption{
  A 3-way even-odd partitioning of function~$f$. Solid lines are observed;
  dashed lines are not.  Note that partitioning describes an organizational
  structure for computing~$f$, but not \emph{how} these computations are
  realized.  In our protocols, $f_1,f_3$ will be computed within an SGX enclave,
  and $f_2$ %will be computed 
  via garbling schemes and oblivious transfer. }
\label{fig:even-odd}
\end{figure}

\subsection{$\ell$-way even-odd partitioning schemes}
Let~$\ell$ be a positive integer and~$f$ be a function of two inputs and two
outputs.
An \emph{$\ell$-way even-odd partitioning scheme}~$\splitter$ for~$f$ is a
sequence of round functions $(f_1, \ldots, f_\ell)$ and a pair of
probabilistic algorithms $(\splitl, \splitr)$.
Let $(a,b) \in \dom f$ and $k$ be a positive integer.  On input~$(1^k, a)$,
algorithm~$\splitl$ outputs an $\ell$-vector of strings~$\veca$.  Similarly, on
input~$(1^k, b)$, algorithm~$\splitr$ outputs an $\ell$-vector of
strings~$\vecb$. These are the local inputs of $\Alice$ and $\Bob$, respectively.
Even-round function evaluations are stateless, while odd-round function
evaluations carry state from one odd-round to the next. This allows us to model
stateful SGX computation.
Even-numbered functions map two strings ($\Alice$ and $\Bob$'s local inputs) to
two strings ($\Alice$ and $\Bob$'s outputs), and
odd-numbered functions map three strings ($\Alice$ and $\Bob$'s local inputs,
and the state of the oracle) to three strings ($\Alice$ and $\Bob$'s outputs and
the oracle's updated state).

\begin{figure}
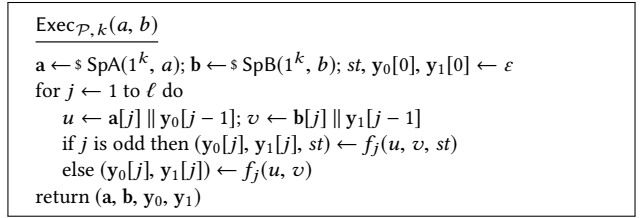

  \oneCol{0.43}
  {
    \underline{$\exec_{\splitter,k}(a,b)$}\\[2pt]
      $\veca \getsr \splitl(1^k, a)$;
      $\vecb \getsr \splitr(1^k, b)$;
      $\st, \vecy_0[0], \vecy_1[0] \gets \emptystr$\\
      \Foreach{j}{1}{\ell}\\
      \ind $u \gets \veca[j] \cat \vecy_0[j-1]$; $v \gets \vecb[j] \cat
      \vecy_1[j-1]$\\
      \ind if $j$ is odd then
        $(\vecy_0[j], \vecy_1[j], \st) \gets f_j(u, v, \st)$\\
      \ind else $(\vecy_0[j], \vecy_1[j]) \gets f_j(u, v)$\\
      return $(\veca, \vecb, \vecy_0, \vecy_1)$
  }
  \caption{Execution of partitioning scheme~$\splitter$ on input $(a,b)$. Note
  that prior outputs are concatenated to the inputs of the next round function.}
  \label{fig:splitter-exec}
\end{figure}

We define the execution of~$\splitter$ in Figure~\ref{fig:splitter-exec}. (See
Figure~\ref{fig:even-odd} for an illustration.)
The partitioning scheme~$\splitter$ is correct for~$f$ if for every positive
integer~$k$ and every $(a,b) \in \dom f$, it holds that
\[
  \Pr\,\big[(\veca, \vecb, \vecy_0, \vecy_1) \getsr \exec_{\splitter,k}(a,b) :
        (\vecy_0[\ell],\vecy_1[\ell]) = f(a,b)\,\big] = 1.
\]
%
% Identity partition
We define~$\idsplitter$ as the 1-way even-odd partitioning scheme defined for
every function~$f$ as follows. Let $(a,b) \in \dom f$ and~$k$ be a positive
integer.  On input $(1^k,a)$, algorithm $\idsplitter.\splitl$ outputs a 1-vector
containing~$a$, and on input $(1^k,b)$, algorithm $\idsplitter.\splitr$ outputs
a 1-vector containing~$b$. Correctness demands that $\idsplitter.f_1 = f$.
We call~$\idsplitter$ the \emph{identity partition}.

\subsection{Partition-relative (2P)-SFE}
Let~$f$ be a function of two inputs and two outputs and let~$\splitter$ be a
partitioning scheme for~$f$. Syntactically, a
protocol~$\proto$ for evaluating~$f$ relative to~$\splitter$ is an oracle
protocol played by $\Alice$ and $\Bob$ with the following correctness condition:
for each $i \in \{\Alice,\Bob\}$, every positive integer~$k$, and every $(a,b)
\in \dom f$, where $(y_\Alice, y_\Bob) = f(a,b)$, it holds that
\begin{equation*}
  \begin{aligned}
    \Pr\Big[ &\veca \getsr \splitter.\splitl(1^k, a);\hspace*{2pt}
              \vecb \getsr \splitter.\splitr(1^k, b):\\
             &\Out_{\proto,k}^i(\veca, \vecb, \emptystr) \outputs y_i\Big] =
             1\,,
  \end{aligned}
\end{equation*}

\paragraph{Defining \sfe}
Our base notion of SFE will be stated relative to a given partitioning
scheme and admits the traditional notion of SFE as a special case.
We formalize the idea that a protocol securely evaluates a function (relative
to a partitioning of that function) if nothing is leaked to either party beyond
that which follows from its own local inputs and outputs.\footnote{%
If $\Alice$'s or $\Bob$'s private input is leaked in the intermediate values,
this is a failure of the partitioning, not of the SFE protocol.}
To do so, we insist that each party's \emph{view} of the protocol's execution
can be computed \emph{without interacting with the other party}.

Let~$f$ be a function, $\splitter$ be a partitioning scheme for~$f$,
and~$\proto$ be an oracle protocol for evaluating~$f$ relative to~$\splitter$.
Security is captured by the game defined in 
%\begin{conference}
%the top-left panel of
%Figure 9 (located in the Appendix of our full paper~\cite{this_full}).
%\end{conference}
%\begin{techreport}
the top-left panel of
Figure~\ref{fig:sfe-notion} (located in the Appendix).
%\end{techreport}
We sketch the notion
here. The security experiment is associated to player~$i\in\{\Alice,\Bob\}$,
adversary~$\advA$, simulator~$\advS$, and security parameter~$k$.
The goal of the adversary is to distinguish the view of player~$i$ from the
output of~$\advS$, which is given~$f$, player~$i$'s local inputs and outputs,
and only the \emph{length} of the other player's inputs and outputs.
To begin, the adversary chooses a pair of inputs $(a,b) \in \dom f$ and the
inputs are split according to~$\splitter$. A challenge bit~$c$ is chosen.
If $c=1$, then the protocol is executed and the adversary is handed~$i$'s view
of the protocol execution, derived from the transcript; if $c=0$, then the
simulator is executed and the adversary is given its output. The adversary wins
if it outputs~$c$.
The advantage of~$\advA$ in the game instantiated with simulator~$\advS$ at
security parameter~$k$ is defined as
\[
  \Adv{\sfe}_{\proto,\splitter,f,i}(\advA, \advS, k) =
   2\cdot \Prob{\Exp{\sfe}_{\proto,\splitter,f,i}(\advA, \advS, k) \outputs \true} - 1.
\]
We say that $\proto$ \emph{securely evaluates}~$f$ \emph{relative
to}~$\splitter$ if for each $i \in \{\Alice, \Bob\}$ and every
polynomial-time~$\advA$, there exists a polynomial-time~$\advS$ such that
$\Adv{\sfe}_{\proto,\splitter,f,i}(\advA, \advS, k)$ is negligible as
a function of the security parameter~$k$.
We refer to such a protocol as an \sfe-secure protocol for~$f$ relative
to~$\splitter$.

When $\splitter = \idsplitter$, our notion reduces to standard SFE,
and we simply say that~$\proto$ \emph{securely evaluates}~$f$,
dropping the reference to the partitioning.
\if{0}
  We specify standard realizations of standard 2P-SFE using a projective
  garbling scheme and an 1-2 OT protocol in Appendix~\ref{sec:gc-based-sfe} and
  using an SGX module in Appendix~\ref{sec:sgx-based-sfe}.
\fi

\section{Hybrid SFE-SGX}\label{sec:hybrid}
Now having discussed our model for Secure Function Evaluation (SFE) on a
partitioned function, we extend this to the context of computing using Intel SGX.
It is critical that we define our own syntax and construction,
despite formalisms already existing~\cite{smcsgx-other,sinha2015moat}, as there is
no previous work that partitions Secure Multiparty Computation 
between SGX hardware and traditional Garbled Circuits.

\vspace{1em}
Let $\splitter$ be an $\ell$-way even-odd partitioning of~$f$, where
$f_1,\ldots,f_\ell$ are the round functions.
\if{0}
  The partition is asymmetric, in the sense that odd-round functions take three
  inputs --- one from~$\Alice$, one from~$\Bob$, and some state --- while
  even-round functions only take the parties' inputs.  Intuitively, the
  odd-round functions are capturing computations within an SGX enclave, which
  may carry state across its executions, and even-indexed functions capture
  non-enclave (and stateless) computations.
\fi
The main technical result of this section says, loosely, that if one possesses
protocols for securely evaluating~$f_1,\ldots,f_\ell$, then these can be
composed to give a protocol for securely evaluating~$f$ \emph{relative to the
information leaked by the component functions.}  This result highlights the need
to carefully consider how~$f$, and the private inputs~$a,b$, are partitioned.
Concretely, one may have a protocol for securely computing
$f_1(a_1,b_1,\st)$, 
which
%i.e., the protocol 
leaks nothing more about
$a_1,b_1,\st$ than~$f_1$ does itself.  But~$f_1$ may, for example, leak all of~$a_1$.
%Thus 
The claimed implication will hold, but if~$a_1=a$, 
%then 
there is no
security in the classical sense of SFE.\footnote{We note that this particular
leakage is ameliorated if $a_1$ itself leaks no efficiently computable
information about~$a$, e.g., $a_1$ is an encryption of~$a$ under a secret key.}

Before we can state our main result, we define (as technical tools) syntax and
security notions for protocols for even- and odd-round function evaluations.

\paragraph{Even-round protocols}
An even-round protocol~$\proto$ is an oracle protocol played by $\Alice$ and
$\Bob$ for evaluating functions of two inputs and two outputs. We write
$\proto[f]$ to denote the protocol instantiated with a particular function~$f$.
Correctness demands that for each $i\in\{\Alice,\Bob\}$, every $(a,b)\in\dom f$,
and every positive integer~$k$, it holds that $\Pr[\Out^i_{\proto[f],k}(a,b)
\outputs y_i] = 1$ where $(y_\Alice, y_\Bob) = f(a,b)$.
We consider the \sfe~security of even-round protocols relative to the identity
partition.

\paragraph{Odd-round protocols}
An odd-round protocol~$\proto$ is an oracle protocol played by $\Alice$ and
$\Bob$ for evaluating functions of three inputs and three outputs. We write
$\proto[f]$ to denote the protocol instantiated with a particular function~$f$.
Correctness demands that for each $i\in\{\Alice,\Bob\}$, every $(a,b,\st) \in
\dom f$, and every positive integer~$k$, it holds that
$\Pr[\Out^{i}_{\proto[f],k}(a,b,\st) \outputs y_i] = 1$, where $(y_\Alice,
y_\Bob, \st^\prime) = f(a,b, \st)$ for some $\st^\prime \in \bits^*$.

\begin{sloppypar} % to prevent equation from going past column boundary
\paragraph{Defining $\wsfef$}
We introduce a new security notion for odd-round protocols. The main distinction
from standard $\sfe$ is that the adversary specifies the oracle's initial state.
Security of odd-round protocols is defined in the 
%\begin{conference}
%top-right panel of
%Figure 9 (located in the Appendix of our full paper~\cite{this_full}).
%\end{conference}
%\begin{techreport}
top-right panel of
Figure~\ref{fig:wsfef-notion} (located in the Appendix).
%\end{techreport}
We define the advantage of adversary~$\advA$ in
the game instantiated with
simulator~$\advS$ as
\[
  \Adv{\wsfef}_{\proto,f,i}(\advA, \advS, k) =
   2\cdot \Prob{\Exp{\wsfef}_{\proto,f,i}(\advA, \advS, k) \outputs \true} - 1.
\]
We say that~$\proto$ is an \wsfef-secure protocol for~$f$ if for each
$i\in\{\Alice,\Bob\}$ and every polynomial-time adversary~$\advA$, there exists a
polynomial-time simulator~$\advS$ such that the function~
$\Adv{\wsfef}_{\proto,f,i}(\advA, \advS, k)$ is negligible as a function of~$f$.
\end{sloppypar}

\subsection{The \mixx~protocol}
Our main protocol is defined in Figure~\ref{fig:proto-hybrid}.
Let~$f$ be a function of two inputs and two outputs and let~$\splitter$ be an
$\ell$-way even-odd partitioning scheme for~$f$.
The protocol is composed from~$\splitter$, an odd-round protocol~$\proto_\odd$,
and an even-round protocol~$\proto_\even$.
The protocol is defined on $\ell$-vectors~$\veca$ and~$\vecb$ corresponding to
$\Alice$ and $\Bob$'s respective split inputs.
They first execute~$\proto_\odd[\splitter.f_1]$ with private inputs $(\veca[1],
\vecb[1])$ with oracle~$\proto_\odd.\O$. As a result, $\Alice$
gets~$\vecy_0[1]$, $\Bob$ gets~$\vecy_1[1]$, and
the oracle's state gets updated to~$\st$ where $(\vecy_0[1], \vecy_1[1], \st) =
\splitter.f_1(\veca[1], \vecb[1], \emptystr)$.
Next, they execute~$\proto_\even[\splitter.f_2]$ on private inputs $(\veca[2]
\cat \vecy_0[1], \vecb[2] \cat \vecy_1[1])$ and with oracle~$\proto_\even.\O$.
$\Alice$ gets~$\vecy_0[2]$ and $\Bob$ gets~$\vecy_1[2]$ as a result.
$\Alice$ and $\Bob$ continue in this way, alternating between odd-round and
even-round evaluations. Correctness of $\splitter$ ensures that
$(\vecy_0[\ell],\vecy_1[\ell]) = f(a,b)$.
\begin{figure}
  \oneCol{0.42}
  {
    \underline{$\proto_\hyb[\splitter,\proto_\odd,\proto_\even](1^k,\veca,\vecb)$}
    \begin{enumerate}[leftmargin=*]
      \item Let $\st, \vecy_0[0], \vecy_1[0] \gets \emptystr$.
        For each~$j$ from~1 to~$\ell$, do as follows:
        let $u = \veca[j] \cat \vecy_0[j-1]$ and $v = \vecb[j] \cat
        \vecy_1[j-1]$.
        If~$j$ is odd, then execute
        \vspace*{-0.5em}
        \[(\vecy_0[j], \vecy_1[j],\pi, \st) \getsr
        \proto_\odd[\splitter.f_j](1^k, u, v, \st)\,;
        \vspace*{-0.5em}
        \]
        otherwise, execute
        \vspace*{-0.5em}
        \[(\vecy_0[j], \vecy_1[j],\pi, \st^\prime) \getsr
        \proto_\even[\splitter.f_j](1^k, u, v, \emptystr)\,.
        \vspace*{-0.5em}
        \]

      \item $\Alice$ halts with private state~$\vecy_0[\ell]$ and $\Bob$ halts
        with private state~$\vecy_1[\ell]$.
    \end{enumerate}
  }
  \caption{The \mixx~protocol, an oracle protocol for evaluating~$f$ constructed
  from $\ell$-way even-odd partitioning scheme~$\splitter$ for~$f$, odd-round
  protocol $\proto_\odd$, and even-round protocol $\proto_\even$.
  }
  \label{fig:proto-hybrid}
\end{figure}

We instantiate the even-round protocol using standard techniques from GC-based
SFE and specify an example, constructed from a projective garbling scheme and a
1-2 transfer protocol, in Figure~\ref{fig:proto-gc}.
\if{0}
  This is similar to the protocol described in Appendix~\ref{sec:gc-based-sfe},
  except $\Alice$ does not send~$d$ (the means to decode the garbled output) to
  $\Bob$. In fact, $\Bob$ is not able to see the final output.
\fi
This protocol does not make use of an oracle, relying only
on the security of its constituent cryptographic primitives.
\begin{figure}
  \oneCol{0.42}
  {
    \underline{$\proto_\gc[f,\garbler,\proto_\ot](1^k,a,b)$}\\[2pt]
    \ind \underline{$\init(1^k)$}: return $(\emptystr, \emptystr, \emptystr)$
    \begin{enumerate}[leftmargin=*]
      \item $\Alice$ generates the circuit $(F, e, d) \getsr \Gb(1^k, f)$, computes
        $A \gets \Enl(e, a)$, and sends $(F, A)$ to $\Bob$.

      \item Let $e = \tokens$. Execute
        \vspace*{-0.5em}
        \[(\emptystr, B, \pi, \st^\prime) \getsr
        \proto_\ot(1^k, e, b, \emptystr).
        \vspace*{-0.5em}
        \]
        (Note that $B = \Enr(e, b)$.)

      \item $\Bob$ computes $Y \gets \Ev(F, A \cat B)$ and sends $Y$ to
       $\Alice$.

      \item $\Alice$ computes $y \gets \De(d, Y)$.

      \item $\Alice$ halts on~$y$ and $\Bob$ halts on~$\emptystr$.
    \end{enumerate}
    \vspace{2pt}
  }
  \caption{An even-round protocol constructed from a projective garbling
  scheme~$\garbler = (\Gb, \En, \Ev, \De)$ and 1-2 transfer
  protocol~$\proto_\ot$.
  See Appendix~\ref{sec:proto-syntax}
  for the semantics of $\init$.}
  \label{fig:proto-gc}
\end{figure}

Finally, we instantiate the odd-round protocol using the SGX module (modeled as
an oracle) and a symmetric encryption scheme $\symenc = (\kg, \enc, \dec)$.
(SGX uses AES-GCM with a random initialization vector. 
%\begin{conference}
%See Appendix A.4 of our full paper~\cite{this_full}
%for syntax and security notions for symmetric encryption.) 
%\end{conference}
%\begin{techreport}
See Appendix~\ref{sec:symenc} 
for syntax and security notions for symmetric encryption.) 
%\end{techreport}
The protocol is defined in Figure~\ref{fig:proto-sgx}. $\Bob$
shares a key with $\Alice$'s enclave. To evaluate a function~$f$ on~$(a,b)$,
$\Bob$ encrypts~$b$ and sends it to $\Alice$, who then sends the function, her
own input, and $\Bob$'s ciphertext to the SGX module.
The module decrypts 
%$\Bob$'s ciphertext,
$b$,
evaluates the function (which depends on its internal state), then returns the
result to $\Alice$.
\begin{figure}
  \oneCol{0.42}
  {
    \underline{$\proto_\sgx[f,\symenc](1^k,a,b,\st)$}\\[2pt]
    \ind \underline{$\init(1^k)$}: $K \getsr \kg$; return $(\emptystr, K, K)$
    \begin{enumerate}[leftmargin=*]
      \item $\Bob$ computes $c \getsr \enc_K(b)$ and sends~$c$ to
        $\Alice$.
      \item $\Alice$ asks $(f, a,c)$ of~$\O(K, \cdot)$.
      \item On input~$(K,(f,a,c))$ and with current state~$\st$,
        oracle~$\O$ computes $b \gets \dec_K(c)$, $(y, \st) \gets f(a, b,
        \st)$, and returns~$y$ to $\Alice$.
      \item $\Alice$ halts on~$y$ and $\Bob$ halts on~$\emptystr$.
    \end{enumerate}
    \vspace{2pt}
  }
  \caption{A odd-round protocol constructed from a symmetric encryption scheme
  $\symenc = (\kg, \enc, \dec)$ and the SGX module, modeled as an oracle
  queried by $\Alice$. 
  See Appendix~\ref{sec:proto-syntax} for the semantics
  of $\init$.}
  \label{fig:proto-sgx}
\end{figure}

The \mixx~protocol instantiated with~$\proto_\sgx$ and~$\proto_\gc$ is useful
for evaluating functions whereby $\Alice$ gets the result and $\Bob$ gets
nothing.
More precisely, $\proto_\hyb[\splitter,\proto_\sgx,\proto_\gc]$ is
well-defined when~$\splitter$ is an even-odd partitioning of a function~$f$ such
that for every $(a,b) \in \dom f$, it holds that $f(a,b) = (y, \emptystr)$ for
some string~$y$.
This is not always desirable; in some applications, ~$\Bob$ should receive
the final result.
To address this, we specify the \emph{dual} protocol of \mixx, whereby $\Bob$
learns all of the intermediate results, including the final result, and~$\Alice$
learns nothing.

\subsection{The dual \mixx~protocol}\label{sec:proto-hybrid-dual}
Let~$f$ be a function of two outputs and two inputs such that for every~$(a,b)
\in \dom f$, there exists a string~$y$ such that $f(a,b) = (\emptystr, y)$.
Let~$\splitter$ be an even-odd partitioning scheme for~$f$.
Protocol $\proto_\gc^\prime$ is like its dual $\proto_\gc$, except that $\Alice$
also sends $\Bob$ the string~$d$, the means to decode, in step 1. In step 3,
$\Bob$ computes $Y \gets \Ev(F, A\cat B)$ and decodes $y \gets \De(d, Y)$ and
halts on~$y$. $\Alice$ halts on output $\emptystr$.
Protocol $\proto_\sgx^\prime$ is like its dual $\proto_\sgx$, except that in
step 3, after computing $(y, \st) \gets f(a,b,\st)$, the oracle
encrypts~$y$ under~$K$ and returns the ciphertext to~$\Alice$. $\Alice$
transmits the ciphertext to $\Bob$ and halts on $\emptystr$. Finally, $\Bob$
decrypts and halts on~$y$.

Then $\proto_\hyb[\splitter,\proto_\sgx^\prime,\proto_\gc^\prime]$ is,
syntactically, a protocol for evaluating~$f$ relative to~$\splitter$. Unlike its
dual, $\Bob$ receives the final and intermediate outputs and~$\Alice$ receives
nothing.\footnote{Some applications may require both $\Alice$ and $\Bob$
to receive intermediate and final outputs. We note that our 
\mixx~and dual \mixx~protocols would
behave like traditional GC if both parties are given the final output, but
intermediate outputs must be handled more carefully. Intermediate outputs
based from odd-rounds
may leak information about the more-sensitive data
handled by even-rounds. This is a line of inquiry beyond the scope
of this paper, but it is certainly important.}

\if{0}
\begin{figure}
  \oneCol{0.42}
  {
    \underline{$\proto_\gc^\prime[f,\garbler,\proto_\ot](1^k,a,b)$}\\[2pt]
    \ind \underline{$\init(1^k)$}: return $(\emptystr, \emptystr, \emptystr)$
    \begin{enumerate}[leftmargin=*]
      \item $\Alice$ generates the circuit $(F, e, d) \getsr \Gb(1^k, f)$, computes
        $A \gets \Enl(e, a)$, and sends $(F, A, d)$ to $\Bob$.

      \item Let $e = \tokens$. Execute\[(\emptystr, B, \pi, \st^\prime) \getsr
        \proto_\ot(1^k, e, b, \emptystr).\]
        (Note that $B = \Enr(e, b)$.)

      \item $\Bob$ computes $Y \gets \Ev(F, A \cat B)$ and $y \gets \De(d, Y)$.
    \end{enumerate}
    \vspace{2pt}
  }
  \caption{An even-round protocol constructed from a projective garbling
  scheme~$\garbler = (\Gb, \En, \Ev, \De)$ and 1-2 transfer protocol~$\proto_\ot$. The
  syntax for the execution of~$\proto_\ot$ is defined in
  Appendix~\ref{sec:proto-syntax}.
  This is the dual protocol of~$\proto_\gc$ defined in
  Figure~\ref{fig:proto-gc}. (Note that~$\Alice$ now sends~$d$ to $\Bob$.)
  }
  \label{fig:proto-gc-dual}
\end{figure}

\begin{figure}
  \oneCol{0.42}
  {
    \underline{$\proto^\prime_\sgx[f,\symenc](1^k,a,b,\st)$}\\[2pt]
    \ind \underline{$\init(1^k)$}: $K \getsr \kg$; return $(\emptystr, K, K)$
    \begin{enumerate}[leftmargin=*]
      \item $\Bob$ computes $c \getsr \enc_K(b)$ and sends~$c$ to
        $\Alice$.
      \item $\Alice$ asks $(f, a,c)$ of~$\O(K, \cdot)$.
      \item On input~$(K,(f,a,c))$ and with current state~$\st$,
        oracle~$\O$ computes $b \gets \dec_K(c)$, $(y, \st) \gets f(a, b,
        \st)$, $z \getsr \enc_K(y)$, and returns~$z$ to $\Alice$.
      \item $\Alice$ sends~$z$ to $\Bob$.
      \item $\Bob$ computes $y \gets \dec_K(z)$.
    \end{enumerate}
    \vspace{2pt}
  }
  \caption{A odd-round protocol constructed from a symmetric encryption scheme
  $\symenc = (\kg, \enc, \dec)$ and the SGX module, modeled as an oracle
  queried by $\Alice$.
  This is the dual protocol of~$\proto_\sgx$ defined in
  Figure~\ref{fig:proto-sgx}.
  }
  \label{fig:proto-sgx-dual}
\end{figure}
\fi

\subsection{Case Study: Database}
Having exhaustively defined the necessary partitioning scheme in developing a hybrid model for SGX,
we present two case studies to demonstrate real-world use of our protocols.

The first of
these involves querying a database.
$\Alice$ provides a database with~$n$ rows, and $\Bob$ issues~$k$ \textsc{select}
queries against 
it.
%the database.
%
Each query retrieves a single 64-bit entry 
%from the database 
associated with the
index 
%value 
provided as input. The results of the queries are returned to
$\Bob$.\footnote{%
Although the database program we architect resembles
a simple key-value store, it offers an initial demonstration that database retrieval
is possible according to our hybrid scheme. No work had previously attempted to combine a 
database application across SGX and GC. Our database supports both
retrieval and updates to data entries in a manner which corresponds to
GET and SET requests against production databases.}
The goal in this setting is to provide the requested
%database 
entries to $\Bob$
without revealing any queried indices to $\Alice$.
Using our hybrid protocol, 
%$\Bob$'s goal is to 
$\Bob$ would protect his most sensitive
queries using GC-based computation and the less sensitive 
%queries 
ones with
SGX-enabled computation. Such a scenario may be practical in the case of a
database implementing a simple Multilevel Security (MLS) scheme, where data is
encoded at one of two levels (i.e., secret, top secret). This approach would
allow the existence of queries to {\it potentially} (but not necessarily) top
secret data to be obfuscated, without incurring the expense of using garbled
circuits for all accesses.

\paragraph{Hybrid Program}
This program can be evaluated using the dual \mixx~protocol.
Some portion of the queries entered by $\Bob$ are considered highly sensitive,
meaning $\Bob$ is especially concerned with not having them revealed.  $\Bob$
splits his $k$ queries into two bins corresponding to $f_1$, the SGX portion of
the program, and $f_2$, the GC portion.
%of the program. 
$\Alice$ loads the database
into her enclave, which will evaluate $f_1$, and $\Bob$ issues queries against
the enclave from his bin of less-sensitive inputs. Once the requested entries
are returned to $\Bob$, the two parties switch from $f_1$ to $f_2$ by having
$\Alice$'s enclave pass the database into the garbled circuit as $y_1$, the
garbled version of the database.  $\Bob$ garbles the inputs corresponding to his
highly sensitive queries by performing 1-2 OT with $\Alice$. $\Bob$ then
evaluates the garbled circuit and receives the requested
%database 
entries
associated with his
%bin of 
more sensitive inputs. $\Alice$ has no output.

\paragraph{Intuition for Further Rounds}
If the output $\Bob$ receives in the second round above advises the next
set of data queries, the program's evaluation may be extended into further rounds.
At this stage, $\Alice$ does not know $\Bob$'s GC output from the prior round,
but any future queries, if dependent on this intermediate output, may
leak information to $\Alice$ about $\Bob$'s sensitive inputs.
If there are many result sets which could have advised $\Bob$'s next 
set of queries (or if $f_2$ was empty),
$\Bob$ may continue as before, splitting his queries into two bins corresponding
to $f_1$ and $f_2$, participating in another series of odd- and even-rounds.
Otherwise (if $\Bob$'s next set of queries could only be a result of a few specific
outputs from the initial GC round), all subsequent queries must be made
exclusively in even-rounds (by emptying $f_1$ or filling it with dummy queries).

\begin{figure}[t]
\centering
\includegraphics[width=2.9in]{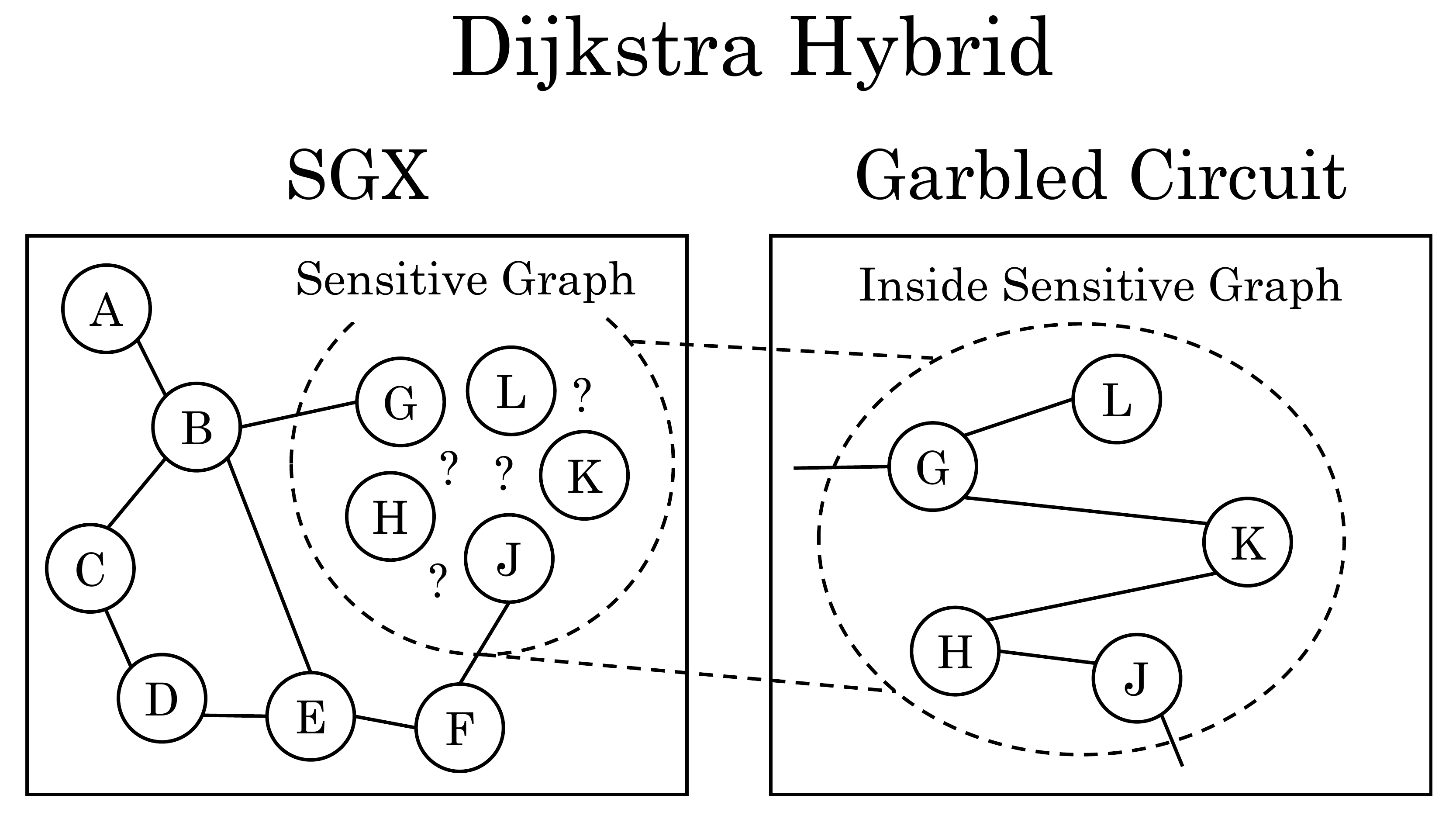}
\caption{A hybrid Dijkstra graph is partitioned into two parts. The majority of the shortest
path is found by the \mixx~computation; the sensitive part of the route is computed using
garbled circuits.}
\label{fig:dijkstra}
\end{figure}

\subsection{Case Study: Dijkstra's Shortest-Path}
Our second case study uses Dijkstra's shortest-path
algorithm. This algorithm finds the shortest path between two nodes in a graph.
Again, there are two parties, $\Alice$ and $\Bob$. The graph topology is known
to both $\Alice$ and $\Bob$, but only $\Bob$ knows the edge weights.  $\Bob$
provides as input the graph's edge weights, and $\Alice$ inputs the starting and
ending points.  The goal of SFE in this setting is for $\Alice$ to receive the
shortest path from her starting point to her ending point while not learning
$\Bob$'s edge weights.

\paragraph{Hybrid Program}
This program is evaluated using the standard \mixx~protocol. Some subset of the
nodes in the graph make up a highly-sensitive portion of the map (graph), and
$\Bob$ is especially concerned with not having the edge weights incident to them
revealed while routing through that portion of the map. The highly-sensitive
portion will be routed solely in a garbled circuit during $f_2$, 
%the GC portion
%of the program, 
whereas the less-sensitive portion will be routed within the SGX module during $f_1$.
%the SGX portion of the program. 
The final route may contain nodes from both
portions 
%of the map 
(Figure~\ref{fig:dijkstra} demonstrates how a graph may
contain both highly-sensitive and less-sensitive portions).  Any route through
the graph is allowed to travel into and out of the sensitive part of the graph a
single time; this is possible by setting edge weights in a specific way.  The
path must not start or end in the highly sensitive part of the graph. This
partitioning could be useful in a scenario where roads traverse private/government
property with selective access, the topological features of which must remain
undisclosed.

The initial Dijkstra computation is run as $f_1$ in $\Alice$'s enclave. $\Alice$'s
input to $f_1$ is the starting and ending points, and $\Bob$ passes the less-%
sensitive edge weights into the enclave. 
%Upon receiving both inputs, $f_1$ is
%evaluated, and a 
A route through the non-sensitive portion of the graph is
calculated.  This route is returned to $\Alice$ from the enclave, though it only
covers the non-sensitive portion of the graph and may not yet be the complete
route. 
%After the initial computation run in the enclave,
%After the initial Dijkstra computation is run in the enclave, 
%the program
Afterward, a garbled circuit is used to
%uses a garbled circuit to 
route through the highly-sensitive portion of the
graph; the garbled circuit is evaluated regardless of whether the path goes
through the highly-sensitive portion of the graph.
%Should the path not need to travel through this more sensitive portion, the
%$f_2$ GC run must still take place in order not to reveal this characteristic
%of the path $f_1$ computed.
The entrance and exit nodes for the sensitive portion of the graph are provided
by the enclave as $y_1$ to $f_2$. $\Alice$ has no input into $f_2$.  $\Bob$'s input
into $f_2$ is the collection of sensitive edge weights.  Upon evaluation of the
garbled circuit associated with $f_2$, $\Alice$ receives as output the route
through the highly sensitive portion of the graph. $\Bob$ has no output.

\paragraph{Intuition for Further Rounds}
At the completion of the first series of rounds, $\Alice$ knows a shortest path between
her starting and ending points. This may be part of a larger path, in the case that
there are certain nodes $\Alice$ must visit en-route. However, $\Alice$'s new starting
and ending points will not be dependent on the prior output.
More rounds may nevertheless be required if, for instance, the same node should not
be revisited, in which case the modified graph (with
visited nodes removed) will be fed directly into the next series of rounds while 
$\Alice$ provides her new starting and ending points (for continuing the path).

\subsection{Security of \mixx}
Our main result is that the composition of an odd-round and an even-round
protocol achieves secure function evaluation relative to a particular
partitioning of the function.
The following theorem says that as long as the evaluations of even-round
functions are \sfe-secure and the evaluations of odd-round functions are
\wsfef-secure, then the \mixx~protocol securely evaluates function~$f$ relative
to even-odd partitioning scheme~$\splitter$ for~$f$.
\begin{conference}
We discuss proofs of this section's theorems in %Appendix A.5 of
our full paper~\cite{this_full}.
\end{conference}
\begin{techreport}
We discuss proofs of this section's theorems in Appendices~\ref{pf:thm-hyb},
\ref{pf:thm-gc}, and \ref{pf:thm-sgx}.
\end{techreport}
\begin{theorem}\label{thm:hyb}
  Let~$f$ be a function of two inputs and two outputs and~$\splitter$ be an
  $\ell$-way even-odd partitioning scheme for~$f$, where $\ell$ is a positive
  integer.
  Let $\proto_\even$ be an even-round protocol and
  $\proto_\odd$ be an odd-round protocol.
  Let $\proto = \proto_\hyb[\splitter,\proto_\odd,\proto_\even]$ as defined in
  Figure~\ref{fig:proto-hybrid}.
  If~$\proto_\odd[\splitter.f_j]$ is an \wsfef-secure protocol for
  evaluating~$\splitter.f_j$ for every odd~$j$
  and~$\proto_\even[\splitter.f_j]$ is an \sfe-secure protocol for
  evaluating~$\splitter.f_j$ for every even~$j$, then $\proto$ securely
  evaluates~$f$ relative to~$\splitter$.
\end{theorem}

The following theorem is a standard result, which says that the composition of a
private, projective garbling scheme and a 1-2 oblivious transfer protocol yields
secure evaluation of the even-round functions.
%We sketch the proof in Appendix~\ref{pf:thm-gc}.
\begin{theorem}\label{thm:gc}
  Let~$f$ be a function of two inputs and two outputs,~$\garbler$ be a projective
  garbling scheme,~$\proto_\ot$ be a 1-2 transfer protocol, and
  let~$\proto=\proto_\gc[f,\garbler,\proto_\ot]$ as defined in
  Figure~\ref{fig:proto-gc}.
  If~$\garbler$ is private and~$\proto_\ot$ is oblivious, then $\proto$ is an
  \sfe-secure protocol for~$f$.
\end{theorem}
This result represents the best-case scenario in the setting where $\Alice$ is
given the internal state of her SGX module. Since even-round evaluations remain
secure, she learns nothing beyond that which follows from the result of each
even-round computation and possession of 
%the private inputs that $\Bob$ sent
the inputs $\Bob$ sent
encrypted to the SGX module.

Lastly, the following theorem says that, viewing the SGX module as an oracle,
secure symmetric encryption suffices for \wsfef-secure evaluation of the
odd-round functions.
($\inddollar$ refers to the standard indistinguishability notion for
encryption schemes 
%\begin{conference}
%defined in Appendix A.4 of our full paper~\cite{this_full}.)
%\end{conference}
%\begin{techreport}
defined in Appendix~\ref{sec:symenc}.)
%\end{techreport}
\begin{theorem}\label{thm:sgx}
  Let~$f$ be a function of three inputs and three outputs, $\symenc = (\kg, \enc,
  \dec)$ be a symmetric encryption scheme, and let~$\proto = \proto[f,\symenc]$
  as defined in Figure~\ref{fig:proto-sgx}.
  If~$\symenc$ is \inddollar-secure, then~$\proto$ is $\wsfef$-secure for~$f$.
\end{theorem}

We remark that the protocol~$\proto_\hyb$ instantiated with~$\proto^\prime_\gc$
and~$\proto^\prime_\sgx$ as the even- and odd-round protocols is also secure.
The proof follows closely the justifications of Theorems~\ref{thm:gc}
and~\ref{thm:sgx}.

In the next section, we justify modeling SGX as a black-box by addressing
side-channel attacks.
But first, we remark that even if these protections do not suffice, our protocol
still provides a measure of assurance.
Suppose, in the worst case, that  $\Alice$ is given the internal state of her
enclave (and hence any keys shared with $\Bob$). Secure evaluation of $f(a,b)$
is out of reach in this setting, since $\Alice$ learns some of $\Bob$'s local
inputs and outputs.
Still, $\Bob$'s most sensitive inputs remain secure
(even upon enclave compromise, 
as an enclave only handles private inputs for odd-rounds): 
$\Alice$ learns nothing
about them beyond that which follows from knowing the state of her enclave and
the result of the computation.

%\input{protocol}
%\input{security}
%%%%%%%%%%%%%%%%%%%%%%%%%%%%%%%%%%%%%%%%%%%%%%%%%%
\section{Side-Channel Attack Mitigations}
\label{sec:side-channel}

Unfortunately, SGX enclaves suffer from side-channels which may leak data regarding
program execution and data~\cite{intel-sgx-enclave-manual-windows,tsgx17,stancesc}.
In order to continue evaluating our hybrid scheme, we
must justify the oracle assumption on SGX heuristically by 
using code modifications to close known side channels.\footnote{%
%closing known side
%channels through the use of code modifications.\footnote{%
For this work, our mitigations are applied manually for each function.
The availability of a tool that automates this process would make
it easier to adopt our hybrid protocol.}
These channels are associated with \emph{timing} and \emph{program flow}, which
allow an outside observer to infer 
%information on 
what path a program takes
during its execution or what memory is being accessed (due to access times),
and~\emph{memory accesses}, where an attacker observes the parts of RAM accessed by
the program.

\paragraph{Timing}
We take steps to ensure our programs run for the same amount of time regardless
of the input, underlying values in memory, or the result. We ensure both
branches of every {\tt if} statement take the same amount of time and fix
loop bounds that could otherwise reveal how many times a loop 
executes~\cite{timingprotections}. An
example of this would be to have each branch perform the same amount of work
(e.g., a single assignment to a variable).

\paragraph{Program Flow}
Even though {\tt if} statements take the same amount of time, controlled-channel
attacks \cite{xu2015controlled} on the program flow are still possible (i.e., if
the \texttt{true} code branch causes a instruction page miss, then this leaks
information).  To further mitigate these, we modify all {\tt if} statements to
prevent branches. For example, 
%the statement 
{\tt if(a == b) {c = 1} else {c =
2}} can be converted into the branch-free version {\tt int t = a == b; t =
makeSameAsBit0(t), c = (t \& 1) | (($\neg$t) \& 2)}, where
\texttt{makeSameAsBit0} sets all bits of \texttt{t} to the least-significant
bit.  
%Notice that, in the new version of the {\tt if} statement,
In the new version, 
there is no
branch that would reveal information about the program flow.\footnote{%
Our approach differs from Raccoon~\cite{raccoon},
which executes multiple
program paths on a given input; instead, we do away
with branching paths altogether prior to execution.}

\label{sec:controlchannelmitigations}

\begin{figure}[t]
\centering
\includegraphics[width=\linewidth]{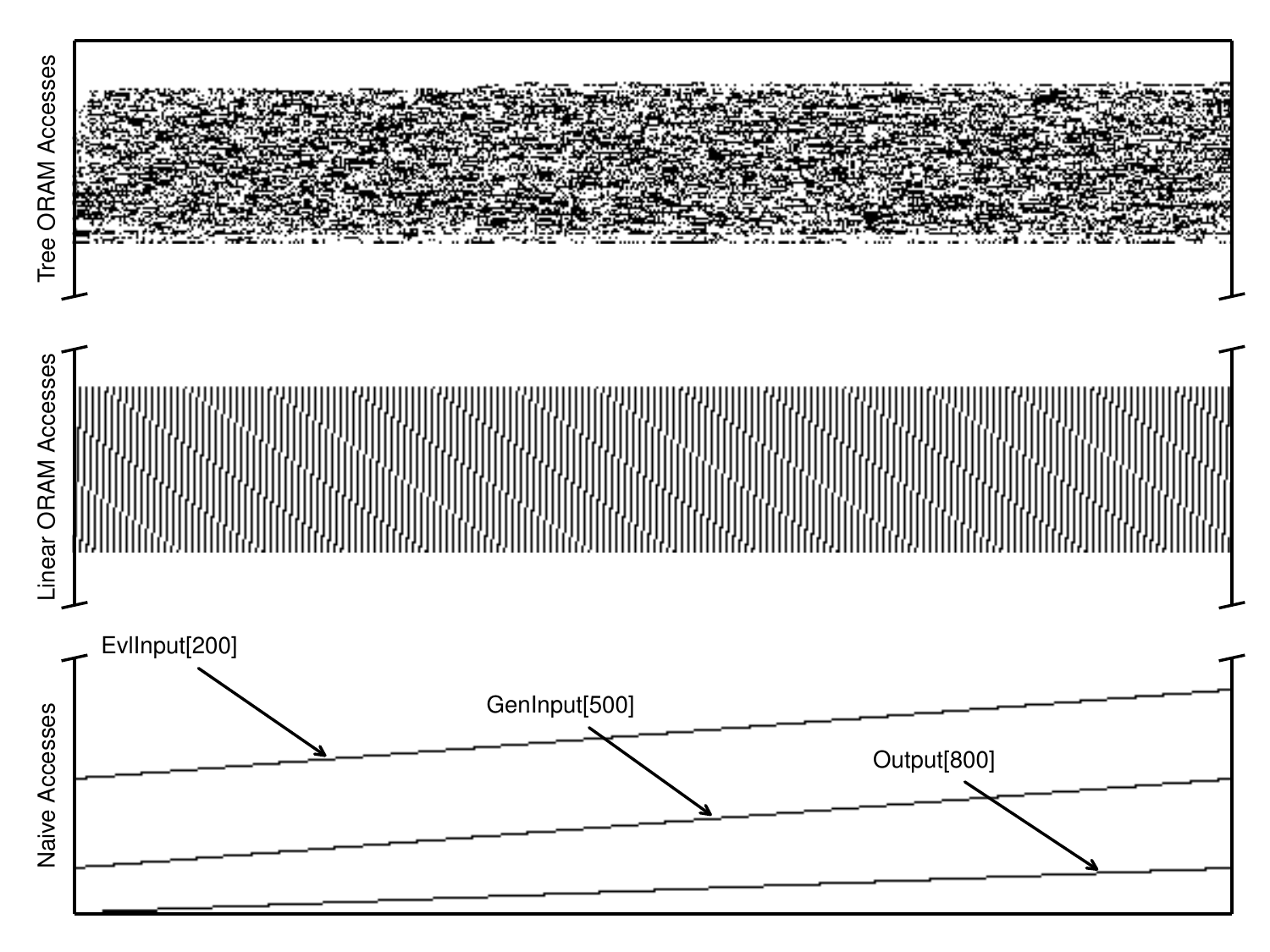}
\caption{A display of memory accesses of the database program. Top: tree-based ORAM; middle: linear search. The three lines on the bottom are unblinded memory accesses. We queried the memory in order from index 0 to $N$.}
\label{fig:oram-graph}
\end{figure}

\paragraph{Memory Accesses}
Any query to a specific array index will reveal the memory page that was queried
if that page faulted.  Additionally, cache side channels can reveal to motivated
attackers which data or code memory addresses are being accessed
\cite{brumley2009cache}.  To counter this, we move RAM accesses from programs to
a binary search tree (BST)-based Oblivious RAM (ORAM)-like construction loosely
based on Path ORAM~\cite{stefanov2013path}.  This modification distributes array
accesses across many randomized memory accesses, as is visually apparent in
\autoref{fig:oram-graph}.
The security model for our BST structure is similar to that of other ORAM
systems, but is different in two important ways: (1) we
deal with a trusted module in a compromised, malicious operating
system~\cite{ren2013design} which may attempt timing-based attacks, and (2) our
modifications are designed to hide program flow, not just data access patterns.\footnote{%
As opposed to traditional (e.g., Path) ORAM, we do not deal with a client/server
setting, in which the server is not trusted to correctly perform the ORAM.
The ORAM functionality within the enclave can be verified during
attestation, making it unnecessary to maintain the position map and stash in a 
separate ``trusted client.''}
%outside the enclave, 
%, as the goal is not to protect
%our accesses from the enclave, but from the system hosting the enclave.}

%\footnote{The
%adversarial model of ORAM is similar to ours. See
%Section~\ref{sec:security} for further details.},
%Our construction
%starts with a standard BST and is

Let us briefly summarize the design of our system.
\begin{itemize}
  \item Our BST forest contains two trees,
    each with 
    %each will contain 
    about half the
    nodes at any given time and in a sorted order, but %will 
    no
    duplicate entries with the other. Each node has a key-value pair, where the
    key is array index and the value is the data.

  \item Every query for a node must descend through each tree all the way to a
    leaf. After the query, each node along the path is removed, moved to a
    different RAM location, and 
    %then 
    randomly added back into a tree. 
    To
    %In order to
    %successfully
    hide where each key and data were moved, we place every
    deleted node's data to a contiguous segment of memory, then randomly swap
    them using the
    const-time and branch-free operations 
    %mentioned 
    from before.
    %our solution to program-flow side-channel attacks.
    %
    %\chris{``using our solution to program-flow side-channel attacks'' is not
    %very clear. What is it that your doing?}

  \item To hide the actual query, $\log(n)$ queries are issued to each tree for
    every real query.
\end{itemize}

Whenever there is code that may reveal the searched index, we employ 
constant-time and branch-free techniques
from the program flow section
to hide the information.

%A proof of our ORAM construction can be found in later sections.

\begin{comment}
An alternative solution is provided by leakage-resilient
cryptography~\cite{dziembowski2008leakage,boyle2012multiparty}. Instead of
trying to prevent any bits of secret data (keys or private inputs) from
being inferred, it operates with a percentage of bits that may be leaked
without compromising privacy, even if this leakage occurs on every
iteration. These schemes are powerful, but a lack of available implementations
limits their usage.
\end{comment}

%\paragraph{Array Accesses in C++}
%To counteract potential array out-of-bounds issues in C++, we move all array
%accesses that are dependent on user input to a safe array type, such as a
%\texttt{vector}, to ensure we use safe access functions. This in turns also
%limits the possibility of memory disclosures to external parties, which could
%reveal secrets or other internal enclave state.
%
%\chris{What's the difference between this and ``memory access''? It's not clear
%to me what this section adds.}

\section{Design and Implementation}
\label{sec:design}
In Section~\ref{sec:hybrid}, we assumed $\Alice$ possessed an SGX module with
which $\Bob$ shared a secret key. Here, and in our implementation, we assume for
simplicity that both parties possess an SGX module.
The two modules share a key unknown to
either $\Alice$ or $\Bob$, constructed during SGX remote attestation
and used to set up a secure communication channel between the modules.
We refer to the party who executes code in his/her
enclave as the \emph{evaluator}. The other party is called the \emph{sender}.

We implement our protocols using Intel's SGX SDK for Linux \cite{01intelsgx}.
% TODO: we need to attack other paper or say this in a different way
%DONE: daveti
Rather than simply replacing GC computation using SGX~\cite{smcsgx-other},
we provide the first implementation that we are aware of 
combining SGX and GC together to perform 2P-SFE,
which is essentially a general form of computation model for 2P-SFE to use SGX.
% SMC using SGX while considering hybrid model. Not just all SGX. We are the middle ground
%We provide one of the first implementations
%
%\chris{of what?}
%which carefully composes SGX and
%garbled circuits to perform 2P-SFE, and we
%
% TODO: sgx hardware isn't that new dude
%that we are aware of on SGX-enabled hardware and 
Unlike previous work~\cite{smcsgx-other},
we also share our experience of using Intel SDK with SGX-enabled hardware to 
develop real-world SFE applications.
% in real world.
%describe
%how the development of real-world SGX applications differs from standard
%development and from simulated or emulated environments.  Previous work, such as
%Haven~\cite{baumann2014shielding} and VC3~\cite{scVC3}
%
%\chris{Do these do SFE?}
%
%used simulators, while the OpenSGX emulator~\cite{jain2016opensgx} implemented
%its own development environment.

\paragraph{Program Partitions}
When developing programs for use in the SGX environment, it is particularly
important to understand the security considerations of code and data use. The
first step
%in developing a program in the SGX environment 
is to partition the
program into two parts. The {\em trusted} partition of the program is kept
within the enclave, while the remainder
%of the application 
is developed outside
of the enclave. This program partitioning is required to minimize the
application's trusted computing base and to save enclave memory. In our
experience, the enclave's memory is fixed by a BIOS setting.\footnote{In our
environment, enclave memory was limited to 128 MB.} This represents a
substantially different programming paradigm from what is required in other
SGX-like environments such as the OpenSGX~\cite{jain2016opensgx} emulator, where programs of any size
are assumed able to fit entirely within the enclave environment.
%are assumed to be able fit entirely within the enclave environment, and no
%partitioning done.
%of the program is performed.

%In our implementation, we used an enclave on both the sender and evaluator for
%simplicity. 
%The shared secret, or key, (used to set up a secure communication channel) is
%constructed in both parties' enclaves during SGX remote attestation.
%To transmit the sender's input, we pass it to the sender's enclave before
%encrypting it under the shared key. Similarly, the evaluator encrypts the
%sender's output in its enclave before returning it to the sender.
% This is said in Section~\ref{sec:hybrid}.
\if{0}
  We use 128-bit AES in GCM mode for encrypting all communications.
\fi
All of our SGX test programs are executed inside of the evaluator's enclave. All
other components, such as code to handle sockets and message processing, are
kept outside the enclave.  In total, $10,172$ SLOC\footnote{Source Lines of Code
(SLOC) generated using David A. Wheeler's `SLOCCount.'} were written between the
sender and evaluator enclaves and their accompanying untrusted applications.
%We implemented the sender and evaluator using FSMs.

\paragraph{Enclave Restrictions}
\if{0} % The heading basically says this.
  Another reason to partition the program is the set of restrictions placed on
  code in an enclave.
\fi
Unlike OpenSGX, where enclave implementations can take
advantage of existing libraries, real enclaves not only require library code to
be statically linked, but also use {\it trusted libraries}. Static linking
guarantees each enclave is self-con-tained and needs no extra libraries to be
installed to where it is deployed.  The trusted libraries created by
Intel for enclave programs are crafted to avoid illegal\footnote{%
We borrow this verbiage from the official Intel SGX documentation.}
instructions (e.g., {\tt fprintf}) that will crash enclave programs at runtime.
%. If an
%enclave program tries to use an illegal
%instruction (e.g., {\tt fprintf}) during
%its execution, it will crash at runtime.
Certain functions (e.g., {\tt strcpy})
are excluded, though variants providing similar functionality (e.g., {\tt strncpy})
are often available.
%do
%not exist at all within the libraries, though variants providing similar
%functionality (e.g., {\tt strncpy}) often do.
% (e.g., {\tt strncpy} can be used within the enclave).
Functions that would otherwise access data outside the enclave
(e.g., {\tt fopen}) are also excluded.

The Intel SGX SDK provides trusted C and C++ libraries, as well as other trusted
libraries (e.g., for SGX runtimes and cryptographic operations). 
We follow these restrictions, primarily using a
%We follow the
%restrictions for developing enclave programs. Consequently, we primarily use a
subset of C functions and some basic C++ built-in data structures (e.g., {\tt
vector}).  When needed, we use the SGX SDK trusted libraries for API
replacements or alternate 
%versions of 
functions. For example, 
%instead of
%calling {\tt rand}, which is not available in the trusted libraries, 
we use {\tt
sgx\_read\_rand}, which accesses the hardware random number generator directly,
instead of {\tt rand}.
%Similarly, instead of using OpenSSL/GnuTLS, we are limited to the cryptographic
%APIs within the trusted libraries.
Similarly, we rely on the trusted libraries' cryptographic APIs instead of OpenSSL/GnuTLS.

\paragraph{SGX/SDK Restrictions}
The SGX remote attestation protocol relies on Intel's Enhanced Privacy ID
(EPID)~\cite{Anati13} technology to verify that the remote enclave is running on
an Intel-authenticated, SGX-enabled CPU.
To use EPID, a signed certificate must first be obtained
from a recognized certificate authority
and registered with Intel Development Services.
For the purposes of this work, we do not purchase a certificate for registration
with Intel and do not implement the EPID part of remote attestation.
%However, at the time of this
%submission,
%
%\chris{Is this still valid for CSF'17}
%yes - daveti
%
%EPID was not available on Linux. 
For our experiments, we instead enable the debug flag
when building our SGX applications to skip EPID. After skipping EPID, the sender
in our setup verifies the measurement of the evaluator's enclave by checking the
signed value contained in the SGX quote (from the attestation) against the known
``good'' quote of the enclave. 
%In summary, for this work, we cannot and do not
%implement the EPID part of the remote attestation.

We encountered instances where we received errors from SGX when attempting to
create and read data from large enclaves.  Unlike normal programs, 
%which do not care about the usage of the stack and heap, 
enclaves require explicit settings
for both
stack and heap size.
% the stack size and heap size. 
Both also need to be 4K aligned (a
normal page size). 
%The default maximum heap size 
%(or total memory available for dynamic allocation) of an enclave is 1 MB.
%%which means that the total memory available for dynamic allocation is 1 MB. 
%Any
%further memory allocation in the enclave would either trigger an out-of-memory
%error if the application is in an ECall, or crash the enclave and application.
For memory-hungry enclaves, 
%one has to tune the enclave configuration file to
%increase the maximum size of stack and/or heap.
maximum stack and/or heap sizes must be increased in the enclave configuration file.\footnote{%
The default maximum heap size 
(or total memory available for dynamic allocation) of an enclave is 1 MB.
%which means that the total memory available for dynamic allocation is 1 MB. 
Any
further allocation would either trigger an out-of-memory
error if the application is in an ECall, or crash the enclave and application.}
%Since enclave memory is reserved in the BIOS, one can foresee a hard limit on
%the maximum size due to a BIOS setting.

\paragraph{Garbled Circuit Implementation}
We used the Frigate semi-honest garbled circuit compiler and semi-honest
protocol implementation provided by Mood et al.~\cite{Mood2016}. This
implementation is a hand-tuned version of the garbled circuit system by Kreuter
et al.~\cite{Kreuter2012}; it uses the point-and-permute~\cite{pointandpermute},
garbled row reduction~\cite{Pinkas2009}, and free XOR~\cite{KS08b} 
optimizations, amongst many others, to improve computational efficiency
and reduce network bandwidth.

%Accordingly, this represents the only
%validated and one of the fastest garbled circuit systems available at
%the time of writing.

\begin{table*}[t]
\centering

\begin{tabular}{l|c|c|c|c|c}
& \multicolumn{5}{c}{Time (ms)} \\
\multicolumn{1}{c|}{Program} & Na{\"i}ve & \multicolumn{2}{c|}{SGX-enabled SFE} & Hybrid & GC\\
\hline
%Millionaires41934304 & X & & &\\\hline
% %UC1000x350x250x250   & & & &\\
% %UC10000x1000x250x250 & & & &\\
% %UC50000x1000x500x500 & X & & &\\
% %UC250000x10000x1500x1500 & X & & &\\
% %UC250000x25000x5000x5000 & X & & &\\\hline
Millionaires1024	& 113 $\pm$ 3\%		& \multicolumn{2}{c|}{114 $\pm$ 2\%}		& -				& 697 $\pm$ 2\%\\
Millionaires4096	& 111 $\pm$ 2\%		& \multicolumn{2}{c|}{110 $\pm$ 2\%}		& -				& 1,640 $\pm$ 1\%\\
Millionaires16384	& 116 $\pm$ 3\%		& \multicolumn{2}{c|}{114 $\pm$ 3\%}		& -				& 5,468 $\pm$ 0.2\%\\
Millionaires262144	& 121 $\pm$ 2\%		& \multicolumn{2}{c|}{121 $\pm$ 2\%}		& -				& 82,960 $\pm$ 0.4\%\\ \hline
Dijkstra20		& 112 $\pm$ 4\%		& \multicolumn{2}{c|}{118 $\pm$ 4\%}		& 1,814 $\pm$ 0.2\%		& 1,086 $\pm$ 0.4\%\\
Dijkstra50		& 111 $\pm$ 3\%		& \multicolumn{2}{c|}{115 $\pm$ 3\%}		& 1,820 $\pm$ 0.2\%		& 4,788 $\pm$ 0.1\%\\
Dijkstra100		& 112 $\pm$ 2\%		& \multicolumn{2}{c|}{120 $\pm$ 3\%}		& 2,333 $\pm$ 0.2\%		& 20,023 $\pm$ 0.02\%\\
Dijkstra200		& 117 $\pm$ 2\%		& \multicolumn{2}{c|}{128 $\pm$ 1\%}		& 6,560 $\pm$ 0.2\%		& 78,905 $\pm$ 0.02\%\\
Dijkstra250		& 119 $\pm$ 2\%		& \multicolumn{2}{c|}{133 $\pm$ 1\%}		& 23,330 $\pm$ 0.2\%		& 122,990 $\pm$ 0.009\%\\
Dijkstra1000		& 125 $\pm$ 2\%		& \multicolumn{2}{c|}{343.9 $\pm$ 0.7\%}	& 51,670 $\pm$ 0.1\%		& 1,972,700 $\pm$ 0.04\%\\
Dijkstra10000		& 412.8 $\pm$ 0.6\%	& \multicolumn{2}{c|}{21,211 $\pm$ 0.03\%}	& 215,940 $\pm$ 0.09\%		& X\\ \hline
			& &Linear 			& Tree & &\\
Database500x2500	& 123 $\pm$ 4\%		 & 150.9 $\pm$ 0.7\%	& 1,663.0 $\pm$ 0.09\% & 18,250 $\pm$ 0.2\%		& 327,390 $\pm$ 0.01\%\\
Database1000x2500	& 116 $\pm$ 3\%		 & 224 $\pm$ 1\%	& 3,218.6 $\pm$ 0.09\% & 45,020 $\pm$ 0.1\%		& 631,200 $\pm$ 0.05\%\\
Database1500x5000	& 117 $\pm$ 1\%		 & 356.2 $\pm$ 0.3\%	& 8,300.1 $\pm$ 0.05\% & 112,800 $\pm$ 0.2\%		& X\\
Database5000x5000	& 116.4 $\pm$ 0.9\%	 & 1,398 $\pm$ 0.1\%	& 31,037 $\pm$ 0.05\%  & 351,800 $\pm$ 0.2\%		& X\\
Database5000x25000	& 146 $\pm$ 1\%		 & 3,538.5 $\pm$ 0.03\%	& 91,919 $\pm$ 0.04\%  & 1,690,000 $\pm$ 0.1\%		& X\\
%Database25000x25000 & X & & &\\\hline

\end{tabular}
\caption{Execution times (in ms) for each protocol, benchmarked with the Unix \texttt{time} command.
All SGX programs were
run for 100 iterations, and GC programs were run for 10 iterations.
%times, and benchmarked with the Unix \texttt{time} command. 
{\tt X}'s
represent runs that would not complete in a timely manner. -'s indicate programs we did not run
due to hybrid Millionaires not being implemented.}
%. The hyphens indicate the hybrid millionaires was not implemented.}
\label{table:timingresults}
\vspace{-2.5em}
\end{table*}

\section{Experiments}
\label{sec:eval}

\begin{comment}
\begin{figure*}[t]
\centering
\includegraphics[width=\linewidth]{figures/eval.png}
\vspace{-3em}
\caption{A bar chart showing the graphical performance results for each program
  and category. The Y axis shows the execution time in milliseconds (in
  $\log$ base-10 scale).  The \texttt{X}'s represent runs that would not complete in
  a timely manner. Please see Table~\ref{table:timingresults} for the confidence
  intervals, which show our evaluation accuracy in terms of percent.
  %
  }
\label{fig:eval}
\vspace{-1em}
\end{figure*}
\end{comment}

We used three common programs in the
SFE literature to evaluate the performance of our implementation.
%to evaluate our \newx~protocol, the \oldx~protocol and
%standard garbled circuits.
We compared
%the performance of 
our \mixx~protocol to \oldx~(function evaluation
in the enclave without side-channel protections), \newx~(with side-channel
protections), and standard \gcx.  
%Our results illustrate that the
%\mixx~protocol enjoys up to a 38-fold improvement over \gcx.

%keep for benjamin - \newx (or \Newx for capitalization) for augmented version,
%\oldx, Oldx for basic sgx for sfe.

We tested our implementation on two HP Envy360 laptops with  Intel quad core
i7-6500U CPUs at 2.50GHz with a 64KB L1 cache, 512KB L2 cache, 4MB L3 cache and
8 GB RAM. Both machines were connected on a VLAN to the same switch via Gigabit
Ethernet.

\subsection{Test Programs}

The two programs used for evaluation of our \mixx~protocol were introduced in
Section~\ref{sec:hybrid}, being Database and Dijkstra's shortest-path. 
%In addition to these, 
Additionally, we use the Millionaires Problem for evaluation of our
\oldx~and \newx~protocols.\footnote{%
We use the Millionaires Problem only to illustrate the differences in performance
of our
\oldx/\newx~protocols and pure GC.
We do not implement a hybrid version, but not due to its incompatibility
with hybrid computation.}
We now present configurations of each
test program, with additional detail for hybrid versions of Database
and Dijkstra.\footnote{%
In Section~\ref{sec:hybrid}, we provided intuition for extending the hybrid versions
of Database and Dijkstra into further rounds. That discussion was
meant primarily to demonstrate that SFE
could be achieved for functions requiring more than two rounds.
We keep our
evaluation versions of these applications at two rounds, as this is sufficient
to show the overhead of transitioning between SGX and GC evaluation.}

\paragraph{Database}
%{\bf \Newx} -
%The database 
%Entries are 64-bits in size.  
The database holds 64-bit entries.
%For our database, 
We experiment using
both the tree-based ORAM (described in
Section~\ref{sec:controlchannelmitigations}) and a simple linear search for
comparison (marked as {\it tree} and {\it linear} in
Table~\ref{table:timingresults}).
For example, {\tt Database500x2500}
%has a database of 500 
%data items and
%entries and 
executes 2,500 select\footnote{%
The time-complexity for set and select queries is
the same in this setup, though it requires the addition of an extra 64 bits of
input per set.}
queries
on 500 entries. 
This program demonstrates the time it would take to use a
database in a larger application with our protocols. As many programs set and
modify entries of the database multiple times in the same run, we have more
queries than the size of the database in our tests to explore the efficiency as
the query size increases.

%{\bf \Mixx} -
For our tests, 
we establish 5\% of the queries to be highly-sensitive
and only ever entered into the garbled circuit in even rounds.
%the other 95\% are entered into the SGX enclave.
%, as they are not as sensitive.

\paragraph{Dijkstra}
%{\bf \Newx} -
%We constructed a graph with $n$ nodes, 
The graph contains $n$ nodes, each with 4 edges. Edge weights are 32
bits.
%
%\chris{Are they integers? In what range? Are they randomly generated?}
%
For example, {\tt Dijkstra20} 
considers 20 nodes.
%represents 
%the Dijkstra program
%the case with 1000
%nodes.

%{\bf \Mixx} -
To test this setup, we define
the number of (1)
nodes in the less sensitive portion of the graph,
(2) nodes in the more sensitive portion of the graph,
and (3) entrances and exits from the more sensitive portion of the graph.
Depending on the start and end points, a route may or may not
need to traverse the sensitive portion of the graph.
Even if the route does not,
we evaluate the circuit
to avoid leaking information
%path does not traverse
%this portion, because not doing so would leak information
about the path computed in the enclave.

Test cases are labeled according to the number of nodes in the less sensitive graph.
We run the following tests:
for the 20-node less sensitive graph,
we consider 12-entrances or exits to the sensitive graph
and a 20-node sensitive graph.
We also run similar tests with the following configurations: 50 (12, 20), 100 (22, 25),
200 (32, 50), 250 (42, 100), 250 (42, 100), 1000 (52, 150),
and 10000 (62, 250).
%, where we have 62 entrances and exits
%from the sensitive graph and a sensitive graph size of 250 nodes.

\paragraph{Millionaires}
Inputs are two $n$-bit unsigned integers.
%The Millionaires Problem takes two $n$-bit unsigned
%integers as input. 
Output is a single bit, informing each party
whose input is larger. For example, {\tt
Millionaires1024}
%represents the millionaires program with 1024 bits of
takes 1024 bits of
input from each party.

\subsection{Results}

\noindent
Table~\ref{table:timingresults} presents our results, which we summarize below.

\paragraph{\Mixx~vs. \Gcx}
By only requiring
part of the computation to use a garbled circuit, we can increase performance by
up to 38x versus pure \gcx~in the case of Dijkstra with 1000 nodes.
%233X vs pure garbled circuits in the case of database 500x2500,
The Database application also demonstrated noticeable improvements, 
%over pure garbled circuits, 
being 18x faster in the 500x2500 case.

Although the \mixx~protocol requires more rounds of computation
by virtue of the splitting between even (GC) and odd (SGX) rounds,
it achieves less communication overhead
than pure \gcx.
In odd rounds, the amount of data exchange is much smaller,
with there being no oblivious transfer or relaying back of lengthy
garbled outputs.
The enclave owner may communicate with its enclave at little or
no cost, while the other party may communicate with the enclave
over the secure channel.
This exchange handled in the odd rounds lets us reduce the size of
the garbled circuit transmitted in even rounds.

The increase in
performance is also dependent upon the size of the computation,
determined by each user's requirements for their data.
We expect the improvement would be even more substantial
for {\tt Dijkstra10000} and larger database cases which were not run.

\paragraph{\Mixx~vs. \Newx}
Our \mixx~protocol combines garbled circuits with \newx, which includes
side-channel protections.  All else being the same, 
use of a garbled circuit results in a performance reduction vs.
%having some of the
%computation use a garbled circuit comes with a performance reduction vs.
\newx~by up to 150x in the worst case, Dijkstra with 1000 nodes.

\paragraph{\Newx~vs. \Gcx}
%We find that the
Our
\newx~protocol is up to 5736x faster than 
a pure GC execution of
%garbled circuit execution of
{\tt Dijkstra1000}.
%the same program run under a
%garbled circuit execution system in the case of {\tt Dijkstra1000}. 
In the
worst case, the \newx~program is only 6x faster; this is for the smallest
experiment, a Millionaires program with 1024 bits of input. In
larger programs, our results show drastic improvements compared
to the garbled circuit implementation.

\paragraph{\Newx~vs \Oldx} Our results show, unsurprisingly, that
without memory protections, the \oldx~protocol outperforms the
\newx~protocol for both Dijkstra and Database, due to the
necessity of hiding the memory access pattern in \newx, by
anywhere from 1.04x (for {\tt Dijkstra50}) 
to 630x (for {\tt Database5000x25000} with Tree ORAM).
The runtime for %our other program,
Millionaires is almost the same for both protocols.

\paragraph{ORAM vs Linear Search} Our results show that the
overhead of the tree ORAM we created is not competitive with a simple
linear search for the program sizes we were able to test; this is
somewhat surprising given the difference in the fraction of the database
that must be searched. The speed of the tree-based ORAM was reduced 
(from the typical $O(polylog(n))$ complexity for ORAM schemes) for
several reasons, including increased time to delete each node from the
tree, and time to mix the node data
after it has been removed from the tree ($O(n^2)$-time complexity).
%Without this
%mixing algorithm, a malicious host could note where nodes were re-added
%for large tree where each piece of data will almost always on its cache
%page.
Thus the linear search is faster for an array database due to a number
of factors including the overhead of the ORAM, branch prediction, and
caching.

\section{Related Work} \label{sec:relwork}

\noindent
\textbf{SFE}
Computation on secure data has long been a goal of the theory and systems
communities. Mechanisms such as Yao's garbled circuit protocol~\cite{Yao1982}
provided proof that arbitrary secure computation was possible, but proved 
too inefficient for practical use. More than two decades later,
Fairplay~\cite{Malkhi2004} provided the first practically efficient
implementation of this construction.
%, requiring only simple hash and symmetric
%key operations to securely evaluate an arbitrary function.  
Since then, a
variety of GC-based SFE protocols have
%been developed in the semi-honest~\cite{Brickell2005,
%Kruger2006,Iliev2010,Huang2011, Lindell2000, Malka2011},
been developed in the semi-honest~\cite{Lindell2000, 
Kruger2006,Huang2011},
covert~\cite{Miyaji2010,Damgard2010, Aumann2010} and malicious 
%models~\cite{Lindell2011, Kreuter2012, HKE13, sS13, Lindell13, Mood2014}.
~\cite{Lindell2011, Kreuter2012, sS13} adversarial models.
Combined with other efforts to reduce
%bandwidth~\cite{Kamara2012,CADT13,CMTB2013, CLT2014, Carter2015}, circuit
bandwidth~\cite{Kamara2012,CMTB2013} and circuit
size~\cite{BHKR2013,KMSB13},
%and improve correctness~\cite{Mood2016},
execution times of applications using garbled
circuits for secure computation have dropped by over five orders of magnitude in the last
decade. A range of privacy-preserving versions of applications have
thus been created, including databases~\cite{Crescenzo14}, navigation~\cite{CMTB2013,
WZPM16}, biometrics~\cite{Carter2015, BCF14},
and genomics~\cite{Huang2011}.
%Garbled circuits have been mixed with homomorphic encryption~\cite{tasty} and
%arithmetic/boolean sharing~\cite{aby}; SecreC~\cite{secrec} even allows transformation
%of code to use the most efficient protocol(s).
However, these applications still introduce substantial computational overhead.
% that far exceed those of computation on unencrypted data.
Bahmani et. al~\cite{smcsgx-other} share our goal of
using SGX to achieve secure computation but leave the
entire computation to the enclave. Ohrimenko et. al~\cite{oblivML}
similarly rely entirely on SGX for multi-party machine learning.
%We realize such a na\"{i}ve
%usage of SGX is inadequate and instead
We realize the limitations of such a na\"{i}ve usage of SGX and instead
%We instead 
propose partition-relative SFE that combines SGX with garbled circuits
for stronger assurances.

\noindent
\textbf{SGX}
Most SGX solutions are in favor of putting the whole application or libOS into an enclave,
including Haven~\cite{haven}, SCONE~\cite{scone}, Graphene-SGX~\cite{graphene-sgx}, and Ryoan~\cite{ryoan},
putting full trust into SGX and holding a large TCB inside the enclave.
While Intel SGX SDK and Panoply~\cite{panoply} mandate program partitioning to minimize the TCB,
the assumption is the security guarantee of enclaves would not be compromised.
A number of attacks have been demonstrated against SGX.
AsyncShock~\cite{asyncshock} exploits synchronization bugs in enclaves using a pre-release version of the Intel SGX SDK.
Controlled-channel attacks~\cite{xu2015controlled} use memory access patterns to exfiltrate sensitive information from secure enclaves.
Cache-based side-channel attacks~\cite{malwaregx,brasser2017software} are also shown possible.
Other side-channel vulnerabilities~\cite{microwalk} are also found within the Integrated Performance Primitives (IPP) cryptographic library used by Intel SGX SDK.
Micro-architecture attacks have also proven to work on enclaves; these include Meltdown~\cite{meltdown} and Spectre~\cite{spectre}.
Foreshadow~\cite{foreshadow} attacks extract the attestation key from enclaves thus breaking SGX remote attestation.
All these attacks show that SGX can be compromised.
%although these attacks could be defeated using the Intel SGX SDK~\cite{recentsgx_sca},
%due to the usage of AES-NI.
%AsyncShock~\cite{asyncshock} exploits synchronization bugs in enclaves using a pre-release version of the Intel SGX SDK.
%Countermeasures include T-SGX~\cite{tsgx}, which leverages Intel TSX instructions to hide the enclave page fault from the untrusted operating system, defending against controlled-channel attacks.
%SGX-Shield~\cite{sgx-shield} enables ASLR for enclave memory space, defending the enclave code against exploitations.
%Sanctum~\cite{sanctum} proposes architecture changes for SGX CPUs, eliminating cache-based attacks.
%ROTE~\cite{rote} leverages multiple SGX machines in a distributed environment to provide a monotonic counter,
%defending SGX sealing against rollback attacks.
Countermeasures include T-SGX~\cite{tsgx}, SGX-Shield~\cite{sgx-shield}, ROTE~\cite{rote}, Sanctum~\cite{sanctum}, etc.,
ranging from software enhancement to hardware changes.
While these defenses are useful to secure SGX applications,
our hybrid approach allows SGX to be compromised by leveraging SFE to secure the most sensitive computation.
%SGXFun and SGX-Reencrypt~\cite{sgxcrypto} explore the technical details in Intel SGX SDK,
%with a focus on crypto operations, and AsyncShock~\cite{asyncshock} exploits synchronization bugs in enclaves using a pre-release of Intel SGX SDK.

\noindent
\textbf{Others}
% Begin Outsourced Computation =================================================
Employing the SGX enclave to securely perform some or all of the evaluation of a
function $f$ echoes secure outsourcing.
% What is Outsourced Computation? ==============================================
Chaum and Pedersen first proposed the idea of outsourcing using ``wallets with
observers''~\cite{walletsWithObservers}.
%wherein a third-party can be assured of the correct behavior of an
%untrusted user by issuing a trusted add-on tamperproof unit, or observer, to the
%user's machine~\cite{walletsWithObservers}.
Hohenberger and Lysyanskaya
introduced the untrusted
% potentially malicious,
but much more computationally
powerful worker~\cite{howtoOutsource}.
% from which secret inputs and outputs must be hidden.
%A preprocessing phase divvies up the computation for outsourcing to
%one or more workers. Secure outsourcing
%is achieved only if secret inputs and outputs remain hidden from the workers.
%protected inputs/outputs available to their operating environment must likewise
%remain hidden from the workers.
%The results of outsourced computation should
%also be verifiable. Given an encoded function and inputs, workers compute
%encoded output values. A malicious worker should not be able to return
%an incorrect, or forged, output that properly decodes.
%Taking it a step further,~\cite{gennaroUntrustedWorkers} proposed a verifiable,
%outsourceable computation scheme as one that does not accept incorrect, or forged,
%output from malicious workers; improved schemes are given by~\cite{zaatar, pinocchio, geppetto}.
Follow-on works founded the idea of a verifiable, outsourceable computation scheme~\cite{gennaroUntrustedWorkers}
and further improved on this idea~\cite{zaatar, pinocchio, geppetto}.
%and improved schemes are given by~\cite{zaatar, pinocchio, geppetto}.
%A verifiable, outsourceable computation scheme is
%first defined in~\cite{gennaroUntrustedWorkers} and realized by combining
%Yao's garbled circuits with a fully homomorphic encryption.
%More efficient schemes include:
%Zaatar~\cite{zaatar}
%, which uses Quadratic Arithmetic Programs (QAPS)
%for probabilistically checkable proofs;
%Pinocchio~\cite{pinocchio}
%, which achieves verification using a proof
%just 288 bytes in length;
%and the recently proposed
%Geppetto~\cite{geppetto}.
% which improves efficiency of proof generation
%by use of generalized QAPs (MultiQAPs),
%bounded proof bootstrapping, and energy-saving circuits.
% Similarities/differences vs outsourced computation ===========================
%In our case, two parties may essentially outsource some portion of the secure
%computation to an enclave owned by one of them.
%Computation within the enclave is efficient, since the contents
%are hidden
%without costly garbled circuit operations.
%An alternative view is that the enclave ``outsources'' the sensitive
%portion of the computation to the two parties, seeking added security
%through GC-based SFE.
%We do not require our scheme to be verifiable
%\mixx~protocol to be verifiable
%outsourced computation scheme 
%as in~\cite{gennaroUntrustedWorkers},
%since correctness 
%of the computed results 
%is not a concern in the semi-honest
%adversarial setting.
% End Outsourced Computation ===================================================
Tram\`er et al.~\cite{sealedglass} previously modeled the enclave as a black-box, albeit
in a weaker sense, by keeping only critical functionalities secret
from the host. 
Town Crier~\cite{towncrier} also treats the enclave as a black-box, trusted for
confidentiality and integrity.
From a theoretical standpoint, our notion of partition-relative SFE is closely
related to the study of non-simultaneous SFE initiated by Halevi, Lindell, and
Pinkas \cite{halevi2011}.
%Here a restricted communication model is considered in
%which each client possessing a private input communicates with a central,
%untrusted server. 
Their result justifies our assertion that the notion of partition-relative SFE is
the strongest-possible security objective in our setting.

\section{Conclusion}
\label{sec:conc}

%Secure function evaluation allows mutually distrusting parties to compute the
%result of a function without revealing their inputs to others.
%%
%The mechanics required to guarantee this property center around the use of
%garbled circuits, which while practical, are still too slow for many
%applications.
%%
%We considered combining traditional GC-based SFE with Intel SGX, which protects
%program and data memory from the operating system, to achieve a somewhat limited
%notion of SFE, but at a substantially lower cost.
%%
%To achieve this, we formalized what it means to partition the computation of a
%function~$f$ into a piece evaluated with SGX and another piece evaluated using
%garbled circuits. We defined secure evaluation of a function relative to such a
%partitioning of that function and proved our scheme achieves this notion.
%%
%The proof is respect to a model in which the SGX module is viewed as an oracle
%available to one of the parties in the protocol; we justified this model
%heuristically by closing known side-channels for the platform.
%%
%Our construction is realized with two example problems: privacy-preserving
%queries to a database and our version of Dijkstra's shortest path algorithm,
%which enables privacy preserving navigation through a graph. Our comprehensive
%evaluation of these problems shows that we achieve a 38x speedup over
%traditional garbled circuit methods with our SGX-enabled SFE.

% daveti: v2
Secure function evaluation (SFE) allows mutually distrusting parties to compute the
result of a function without leaking anything to others but suffers from large runtime overhead.
Intel SGX provides a natural environment for secure computation,
but side-channel and micro-architecture attacks can leak data from enclaves.
We propose a hybrid approach to combine SFE and SGX, 
by formalizing what it means to partition the computation of a
function~$f$ into a piece evaluated with SGX and another piece evaluated using
SFE (garbled circuits). We defined secure evaluation of a function relative to such a
partitioning of that function and proved our scheme achieves this notion.
%Our construction is realized with two example problems: privacy-preserving
%queries to a database and our version of Dijkstra's shortest path algorithm,
%which enables privacy preserving navigation through a graph.
Our comprehensive
evaluation of two case studies shows that we achieve a 38x speedup over
traditional garbled circuit methods with our SGX-enabled SFE and points to a way
to ensure safer and faster secure computation.

%\begin{techreport}
%\subsection*{Acknowledgements}
%This work is supported by the US National Science Foundation
%under grant numbers CNS-1540217, CNS-1562485, CNS-1564446, and CNS-1564444.
%\end{techreport}

\begin{acks}
This work is supported in part by  
the~\grantsponsor{1}{US National Science Foundation}{https://www.nsf.gov/} 
under grant numbers~\grantnum{11}{CNS-1540217},~\grantnum{12}{CNS-1564444}, and~\grantnum{12}{CNS-1642973}.
Any opinions, findings, and conclusions or recommendations expressed in this material
are those of the authors and do not necessarily reflect the views of the National Science Foundation.
\end{acks}

\bibliographystyle{abbrv}
\bibliography{references,pt-refs,grant,daveti}

%\begin{techreport}
\appendix
\section{Appendix}
\begin{figure*}[t]
  \twoColsTwoRows{0.42}
  {
    \underline{$\Exp{\sfe}_{\proto,\splitter,f,i}(\advA, \advS, k)$}\\[2pt]
      $c \getsr \bits$\\
      $(a, b, \sigma) \getsr \advA(\pick, 1^k, f)$\\
      if $(a,b) \not\in \dom f$ then return $\bot$\\
      $(L_0, L_1, L_\O) \getsr \proto.\init(1^k)$;
      $(\veca, \vecb, \vecy_0, \vecy_1) \getsr \exec_{\splitter,k}(a,b)$\\
      $(z_0, z_1, \pi, \st^\prime) \getsr \proto(1^k, \veca, \vecb, \emptystr)$\\
      if $c=1$ then $\omega \gets \View^i_{\proto,k}(\pi)$\\
      else if $i=\Alice$ then $\omega \getsr \advS(1^k, \veca, \len{\vecb}, \vecy_0)$\\
      else if $i=\Bob$ then $\omega \getsr \advS(1^k, \len{\veca}, \vecb, \vecy_1)$\\
      $c^\prime \getsr \advA(\guess, \sigma, \omega)$\\
      return $(c = c^\prime)$
  }
  {
    \underline{$\Exp{\wsfef}_{\proto,f,i}(\advA, \advS, k)$}\\[2pt]
      $c \getsr \bits$\\
      $(a, b, \st, \sigma) \getsr \advA(\pick, 1^k, f)$\\
      if $(a,b,\st) \not\in \dom f$ then return $\bot$\\
      $(L_0, L_1, L_\O) \getsr \proto.\init(1^k)$\\
      $(y_0, y_1, \pi, \st^\prime) \getsr \proto(1^k, a, b, \st)$\\
      if $c=1$ then $\omega \gets \View^i_{\proto,k}(\pi)$\\
      else if $i=\Alice$ then $\omega \getsr \advS(1^k, a, |b|, y_0, |\st|)$\\
      else if $i=\Bob$ then $\omega \getsr \advS(1^k, |a|, b, y_1, |\st|)$\\
      $c^\prime \getsr \advA(\guess, \sigma, \omega)$\\
      return $(c = c^\prime)$
  }
  {
    \underline{$\Exp{priv}_{\garbler}(\advA, \advS,k)$}\\[2pt]
      $c \getsr \bits$\\
      $(f, x, \sigma) \getsr \advA(\pick, 1^k)$\\
      if $x \not\in \dom f$ then return $\bot$\\
      if $c=1$ then $(F, e, d) \getsr \Gb(1^k, f)$; $X \gets \En(e, x)$\\
      else $(F, X, d) \getsr \advS(1^k, f, f(x))$\\
      $c^\prime \getsr \advA(\guess, F, X, d, \sigma)$\\
      return $(c=c^\prime)$
  }
  {
    \underline{$\Exp{pfe}_{\proto,i}(\advA, \advS, k)$}\\[2pt]
      $c \getsr \bits$\\
      $(f, x, \sigma) \getsr \advA(\pick, 1^k)$\\
      if $x \not\in \dom f$ then return $\bot$\\
      $(L_0, L_1, L_\O) \getsr \proto.\init(1^k)$\\
      $(y, \emptystr, \pi, \st^\prime) \getsr \proto(1^k, f, x, \emptystr)$\\
      if $c=1$ then $\omega \gets \View_{\proto,k}^i(\pi)$\\
      else if $i = \Alice$ then $\omega \getsr \advS(1^k, f, |x|, \emptystr)$\\
      else if $i = \Bob$ then $\omega \getsr \advS(1^k, |f|, x, f(x))$\\
      $c^\prime \getsr \advA^\O(\guess, \sigma, \omega)$\\
      return $(c=c^\prime)$
  }
  \caption{Security notions for
  function evaluation \textbf{(top-left)},
  odd-round protocols \textbf{(top-right)},
  private garbling schemes \textbf{(bottom-left)},
  and private function evaluation \textbf{(bottom-right)}.
  Let $\proto = (\init, \proc, \O)$ be an oracle protocol and $\garbler = (\Gb,
  \En, \Ev, \De)$ be a garbling scheme. See the Appendix for additional
  notation.}
  \vspace{6pt}
  \hrule
  \label{fig:sfe-notion}
  \label{fig:ot-notion}
  \label{fig:wsfef-notion}
  \label{fig:priv-notion}
\end{figure*}

%%%%%%%%%%%%%%%%%%%%%%%%%%%%%%%%%%%%%%%%%%%%%%%%%%%%%%%%%%%%%%%%%%%%%%%%%%%%%%%
Let $[1..\ell]$ denote the set of integers from~1 to~$\ell$.
If $\veca$ is a vector
over strings, let $\len{\veca}[i] = |\veca[i]|$ for every $i \in [1..|\veca|]$.
A function $\epsilon(\cdot)$ is \emph{negligible} if for every positive
polynomial $p(\cdot)$, there exists a $k_0$ such that for every $k \ge k_0$, it
holds that $\epsilon(k) < 1/p(k)$.
Both an \emph{adversary} and a \emph{simulator} are randomized algorithms.

\subsection{Protocols}
\label{sec:proto-syntax}
We define syntax and execution semantics for two-party protocols in which one or
both players have access to an explicitly defined oracle.
An \emph{oracle-relative, two-party protocol} is a triple
$\proto=(\init, \proc, \O)$ where~$\init$ and~$\proc$ are randomized
algorithms and~$\O$ is an oracle.
Let~0 and~1 denote the (non-oracle) protocol parties.

% Init
Fix security parameter~$k \geq 0$. Algorithm~$\init$ takes as input~$1^k$ and
outputs a triple of strings $(L_0, L_1, L_\O)$ called the \emph{long-term
inputs} of parties~0, 1 and oracle~$\O$ respectively.
This is written $(L_0,L_1,L_\O) \getsr \init(1^k)$.

% Proc
Algorithm~$\proc$ takes as input the identity~$i$ of a message recipient, their
long-term input $L_i$, their current state~$\st_i$, and the message~$m$ being
delivered.
It outputs a message~$m^\prime$ (to be sent to~$1-i$), the
coins~$\coins_i$ used in the computation, the updated state~$\st_i^\prime$,
and an indication of whether to halt $(\bot)$ or continue ($\top$).
The algorithm is given oracle access to~$\O(L_\O,\cdot)$. On
input~$(L_\O, x)$, the oracle performs some operation (specified by the
protocol) on the string~$x$ and its current state, updates its state, and
returns a string to the player.
This is written $(m^\prime,\coins_i,\st_i^\prime,\delta) \getsr
\proc^{\O}_i(L_i,\st_i,m)$ where $\delta \in \{\bot,\top\}$.

% Protocol execution.
Protocols are executed with respect to the players' private inputs.
Executing the protocol with input $(a_0,a_1)$, where $a_0$ and $a_1$ are strings
(or vectors over strings) means to initialize $\st_i=(1^k, a_i)$ for $i \in
\{0,1\}$, and facilitate the exchange of messages between~0 and~1 until both
halt. More formally, the protocol consists of a sequence of calls
to~$\proc_i^\O(\cdot, \cdot, \cdot)$ alternating between $i=0$ and $i=1$. The
protocol specifies which player starts.
This is denoted
\[
  (y_0, y_1, \pi, \st^\prime) \getsr \proto(1^k, a, b, \st).
\]
% Transcript
The transcript consists of the inputs $1^k$, $a_0$, $a_1$, and~$\st$, the
outputs of each call to~$\proc$, and all of the oracle queries made by~$\proc$,
along with their responses.

% Variables associated to protocols.
Let $\omega = \View_{\proto,k}^i(\pi)$ denote the view of player~$i$ for the
particular execution of $\proto$ transcribed by~$\pi$. In particular,
string~$\omega$ encodes~$i$'s initial state, and the portions of~$\pi$ that
correspond to an execution $\proc_i^{\O}(\cdot,\cdot,\cdot)$, i.e. those
corresponding to party~$i$.
\if{0}
  Let $\Out_{\proto,k}^i(a, b, \st)$ be a random variable denoting the final
  state of player~$i$ upon executing the protocol on input $(1^k, a, b, \st)$
  where~$i \in \{\Alice, \Bob\}$.
\fi

\if{0}
  % Point about reductions.
  An adversary may execute a protocol (e.g. as part of a reduction) without having
  access to the other player's or the oracle's long-term input.
  This echoes the fact that we view the adversary as one of the players in the
  protocol.
  It may also initiate a protocol with long-term inputs of its choosing (e.g. as
  part of a reduction).
\fi

\paragraph{Composing oracle protocols}
Let $(\proto_1, \ldots, \proto_\ell)$ be a sequence of protocols. A
protocol~$\proto$ constructed from their composition is defined as follows. The
initialization algorithm $\proto.\init$ executes each $\proto_j.\init$ and
returns the triple $(\vec{L}_0, \vec{L}_1, \vec{L}_\O)$ where $\vec{L}_0[j]$
(resp. $\vec{L}_1[j]$ and $\vec{L}_\O[j]$) denotes the long-term input of
player~0 (resp. player~1 and oracle $\proto_j.\O$) in protocol~$\proto_j$.
Executing $\proto$ on inputs $(1^k, a_0, a_1, \st)$, where~$a_0$ and~$a_1$
are strings (or vectors over strings) means to iteratively execute $\proto_j$
for each~$j$ from~1 to~$\ell$ on inputs specified by the protocol, except that
$\st$ is the initial state of oracle~$\proto_1.\O$.  During the execution of
$\proto_j$, the algorithm $\proto_j.\proc$ is given oracle access to
$\proto_j.\O(\vec{L}_\O[j], \cdot)$.
The output is the tuple $(y_0, y_1, \pi, \st^\prime)$ where $y_0$  and~$y_1$
are the final states of players~0 and~1 respectively after
executing~$\proto_\ell$, string $\pi = (\pi_1, \ldots, \pi_\ell)$, where $\pi_j$
is the transcript of the $j$-th protocol, and~$\st^\prime$ is the final state
of oracle~$\proto_\ell.\O$.
Finally, let $\View_{\proto,k}^i(\pi) =
(\View_{\proto_j,k}^i(\pi_j))_{j=1}^\ell$.

%%%%%%%%%%%%%%%%%%%%%%%%%%%%%%%%%%%%%%%%%%%%%%%%%%%%%%%%%%%%%%%%%%%%%%%%%%%%%%%
\subsection{Garbling schemes}\label{sec:garbler}
A garbling scheme $\garbler = (\Gb, \En, \Ev, \De)$ is \emph{projective}
if~$e$ (the second output of the garbling algorithm) is a sequence $\tokens$ of
strings called \emph{tokens} such that
$\En(e, x) = (X_1^{x_1}, \ldots, X_n^{x_{n}})$ where $x=x_1 \cdots x_n$.

\begin{sloppypar} % to prevent equation from breaching column boundary
Let $\garbler$ be a projective garbling scheme, $f : \bits^n \to \bits^*$ be a
function, and $(F, e, d)$ be a triple of strings in the range of
$\Gb(1^k, f)$.  Let~$\ell \leq n$ be an integer, and let~$r=n-\ell$. Then if $a \in
\bits^\ell,b \in \bits^{r}$, and~$x = a \cat b$, we define $\Enl(e,a)$ and
$\Enr(e,b)$ so that $\En(e,x)=\Enl(e,a)\concat\Enr(e,b)$.  That is,
$\Enl(e, a) = (X_1^{a_1}, \ldots, X_{\ell}^{a_\ell})$ and $\Enr(e, b) =
(X_{\ell+1}^{b_1}, \ldots, X_n^{b_r})$.
\end{sloppypar}

\paragraph{PRIV}
We specify the simulation-based notion of~\cite{garbling} (instantiated in our
setting) in the bottom-left panel of Figure~\ref{fig:priv-notion}.
% Intuition
This game captures an adversary's advantage in distinguishing the output of the
garbling algorithm on input $(f, x)$ from the output of the simulator on input
$(f, f(x))$. A garbling scheme is ``secure'' if for every reasonable adversary,
there exists a simulator such that this advantage is ``small''.  Intuitively,
this captures the idea that if a garbling scheme is secure, then possession of
the garbled function, garbled input, and the final output leaks only a
negligible amount information about the input~$x$ to the circuit evaluator.
We define the advantage of $\advA$ in the game with simulator $\advS$ at
security parameter $k$ as
\[
  \Adv{priv}_{\garbler}(\advA, \advS, k) =
  2 \cdot \Prob{\Exp{priv}_{\garbler}(\advA, \advS,k) \outputs \true} - 1.
\]
We say that~$\garbler$ is \emph{private} if for every polynomial-time
adversary~$\advA$, there exists a polynomial-time simulator~$\advS$ such that
the function $\Adv{priv}_{\garbler}(\advA,\advS,k)$ is a negligible function
of~$k$.

%%%%%%%%%%%%%%%%%%%%%%%%%%%%%%%%%%%%%%%%%%%%%%%%%%%%%%%%%%%%%%%%%%%%%%%%%%%%%%%
\subsection{1-2 oblivious transfer}\label{sec:ot}
A \emph{1-2 transfer protocol} is a two-party protocol~$\proto$ played by
$\Alice$ and $\Bob$ with the following correctness condition:
when executed with $\Bob$'s private input $b = b_1\cdots b_n \in \bits^n$ and
$\Alice$'s private input
of a sequence of tokens $\tokens$, it holds that
\[
  \Prob{\Out^\Bob_{\proto,k}(\tokens, b) \outputs
        (X^{b_1}_1, \ldots, X^{b_n}_n)} = 1 \]
and
\[
  \Prob{\Out^\Alice_{\proto,k}(\tokens, b) \outputs
        \emptystr} = 1 \,,
\]
where $|X_j^0| = |X_j^1|$ for each $j \in [1..n]$.

We define oblivious transfer (OT) in the presence of a semi-honest adversary.
Following \cite[section 7.1]{garbling},
we formulate the security of OT as an instance of \emph{private function
evaluation}, or PFE. Here Alice has private function~$f$ and Bob has a
private input~$x$ in the domain of~$f$. The goal is that Bob learns $f(x)$
without learning anything about~$f$ (except $|f|$) and Alice learns noting
about~$x$ (except~$|x|$).
In the case of 1-2 OT, the private function corresponds to the map $b \mapsto
(X^{b_1}_1, \ldots, X^{b_n}_n)$ (and hence encodes $\Alice$'s tokens) and the
private input is $\Bob$'s string~$b$.
The length of this map is a function of~$|\tokens|$.

\paragraph{PFE}
Security is captured by an experiment defined in the bottom-right of Figure~\ref{fig:ot-notion}
associated to player~$i\in\{\Alice,\Bob\}$, adversary~$\advA$, and
simulator~$\advS$. If $i=\Alice$, the goal of the adversary is to distinguish
$\Alice$'s view from the output of~$\advS$, which is given as input the security
parameter, the function, and the length of $\Bob$'s private input. If $i=\Bob$,
the goal of the adversary is to distinguish player $\Bob$'s view from the output
of~$\advS$, which is given as input the security parameter, the input~$x$, the
value $f(x)$, and the length of the private function.
The advantage of~$\advA$ in the game instantiated with simulator~$\advS$ at
security parameter~$k$ is defined as
\[
   \Adv{pfe}_{\proto,i}(\advA, \advS,k) =
    2\cdot \Prob{\Exp{pfe}_{\proto,i}(\advA, \advS,k) \outputs \true} - 1.
\]
We say that~$\proto$ is PFE-secure if for each $i\in\{\Alice,\Bob\}$ and every
polynomial-time adversary~$\advA$, there exists a polynomial-time
simulator~$\advS$ such that the function $\Adv{pfe}_{\proto,i}(\advA, \advS, k)$
is a negligible function of~$k$.

\paragraph{OT}
We say that a 1-2 transfer protocol is \emph{oblivious} if it is a secure PFE
protocol.

%%%%%%%%%%%%%%%%%%%%%%%%%%%%%%%%%%%%%%%%%%%%%%%%%%%%%%%%%%%%%%%%%%%%%%%%%%%%%%%
\subsection{Symmetric encryption}\label{sec:symenc}
\begin{sloppypar} % to prevent equation from breaching column boundary
We give the standard concrete security notion for symmetric encryption.
A symmetric encryption scheme~$\symenc$ is a triple of randomized algorithms
$(\kg, \enc, \dec)$.
Algorithm~$\kg$ outputs a string~$K$ called the key.
Algorithm~$\enc$ takes as input the key~$K$, a message~$M\in\bits^*$, and
outputs a ciphertext~$C\in\bits^*$.
Algorithm~$\dec$ takes as input the key~$K$, a ciphertext~$C$, and
deterministically outputs the message~$M$.
Correctness demands that for key~$K$ and every message~$M$ (in the implicit
message space), it holds that $\Prob{ C \getsr \enc_K(M): \dec_K(C) = M} = 1$.
\end{sloppypar}

\paragraph{\inddollar}
We associate to an adversary and encryption scheme an experiment in which the
adversary is given an oracle, which it may query any number of times. After
interacting with its oracle, it outputs a bit.
We define the advantage of an adversary~$\advA$ in attacking $\symenc$ as
\[
  \Adv{\inddollar}_\symenc(\advA) =
  \Prob{K \getsr \kg: \advA^{\enc_K(\cdot)} \outputs \true} -
  \Prob{\advA^{\$(\cdot)} \outputs \true}
\]
where oracle~$\$(\cdot)$ outputs a random bit string of appropriate length, i.e. as
long as a ciphertext corresponding to the query would be.

\begin{techreport}
%%%%%%%%%%%%%%%%%%%%%%%%%%%%%%%%%%%%%%%%%%%%%%%%%%%%%%%%%%%%%%%%%%%%%%%%%%%%%%%
\subsection{Proof of Theorem~\ref{thm:hyb}}\label{pf:thm-hyb}
It suffices to prove the following claim:
Let $f_j = \splitter.f_j$ for every $j \in [\ell]$ and let
$i\in\{\Alice,\Bob\}$.
Let~$\advA$ be an adversary and~$\advS_\odd$ and~$\advS_\even$ be simulators.
There exist adversaries~$\advB_\odd$ and~$\advB_\even$ and a simulator~$\advS$
such that
\begin{equation*}
  \begin{aligned}
    \Adv{\sfe}_{\proto,\splitter,f,i}(\advA, \advS, k) \le&
     \,2\sum_{h=1}^{\lceil \ell/2 \rceil}
      \Adv{\wsfef}_{\proto_\odd[f_{2h-1}],f_{2h-1},i}(\advB_\odd, \advS_\odd,
      k)
      \\+&
     \,2\sum_{h=1}^{\lfloor \ell/2 \rfloor}
      \Adv{\sfe}_{\proto_\even[f_{2h}],\idsplitter,f_{2h},i}(\advB_\even, \advS_\even, k).
  \end{aligned}
\end{equation*}
Moreover, if~$\advA$, $\advS_\odd$, and~$\advS_\even$ are polynomial-time, then
so are~$\advB_\odd$, $\advB_\even$, and~$\advS$.

% Odd rounds
We first construct the adversary~$\advB_\odd$ from $\advA$ and the
simulator~$\advS$ from $\advS_\odd$ and $\advS_\even$.
Let $j\in[1..\ell]$ be odd. We sketch the construction of adversary~$\advB_\odd$
on input for odd-round function~$f_j$.
Let~$\advB_\odd$ be the adversary specified in
Figure~\ref{fig:pf-thm-hyb-adv}.
% pick
In its $\pick$ phase, adversary~$\advB_\odd$ is given as input $(1^k, f_j)$.
It first executes~$\advA$ in its $\pick$ phase on input of the security
parameter and~$f$. When $\advA$ outputs $(a, b, \sigma)$ where $(a,b) \in \dom
f$ and $\sigma \in \bits^*$, adversary $\advB_\odd$ splits~$a$ and~$b$ according
to~$\splitter$, then evaluates~$f$ according to~$\splitter$ up to the $(j-1)$-th
round. It returns the appropriate inputs to the $j$-th round as its choice and
specifies $(\sigma, j)$ as its state to carry-over to the next phase of the
game. Let $(u,v,\st, (\sigma, j))$ denote $\advB_\odd$'s output.

% Choose
Let~$d$ denote the challenge bit in the \wsfef~game played by $\advB_\odd$. In
its $\guess$ phase, adversary~$\advB_\odd$ gets as input a string~$\omega$,
which is player~$i$'s view of $\proto_\odd[f_j]$ executed on input $(1^k, u, v,
\st)$ if $d=1$, and the output of the simulator (given player~$i$'s inputs)
otherwise.
Adversary~$\advB_\odd$ flips a coin~$d^\prime$. If it comes up heads, then it
emulates $\advA$ in the \sfe~game when its challenge bit is~1; otherwise it
emulates $\advA$ in the \sfe~game instantiated with a simulator~$\advS$, which
is specified by~$\advB_\odd$. In particular, player~$i$'s view of odd-round
protocol executions is simulated by $\advS_\odd$ and even-round protocol
executions are simulated by $\advS_\even$.

% The reduction.
Finally, adversary~$\advB_\odd$ substitutes the (possibly-simulated) view of the
$j$-th round execution with its own input~$\omega$. It then executes $\advA$'s
$\guess$ phase on input of the emulated view and $\advA$'s carry-over state
$\sigma$. When $\advA$ outputs its guess $c^\prime$, adversary~$\advB_\out$
outputs~$c^\prime$.

\begin{sloppypar} % to prevent equation from breaching column boundary
% Analysis
If $d=1$ and $d^\prime=1$, then the reduction perfectly emulates $\advA$ in the
\sfe~game when its challenge bit is~1; on the other hand, if $d=0$ and
$d^\prime=0$, then the reduction perfectly emulates~$\advA$ instantiated
with~$\advS$ with the challenge bit being~0.
Let $X_j = \Exp{\wsfef}_{\proto_\odd[f_{j}],f_{j},i}(\advB_\odd, \advS_\odd,
k)$.  By construction, we have that
\[
  \Prob{X_j\outputs\true | d=1, d^\prime=1} = \Prob{\Exp{\sfe}_{\proto,\splitter,f,i}(\advA, \advS, k)
  \outputs \true | c=1}
\]
where~$c$ denotes the challenge bit in $\advA$'s game. Similarly, we have
\[
  \Prob{X_j\outputs\true | d=0, d^\prime=0} = \Prob{\Exp{\sfe}_{\proto,\splitter,f,i}(\advA, \advS, k)
  \outputs \true | c=0}.
\]
Hence, and since~$d$ and~$d^\prime$ are independent,
\begin{equation*}
  \begin{aligned}
    \Prob{X_j\outputs \true} &\ge \frac{1}{4}\Big(
      \Prob{\Exp{\sfe}_{\proto,\splitter,f,i}(\advA, \advS, k) \outputs \true | c=1}\\
      &\hspace*{15pt}+ \Prob{\Exp{\sfe}_{\proto,\splitter,f,i}(\advA, \advS, k) \outputs \true | c=0}
    \Big)\\
    &= \frac{1}{2} \cdot \Prob{\Exp{\sfe}_{\proto,\splitter,f,i}(\advA, \advS, k)
    \outputs \true}.
  \end{aligned}
\end{equation*}
It follows that
\[
  2 \cdot \Adv{\wsfef}_{\proto_\odd[f_{j}],f_{j},i}(\advB_\odd, \advS_\odd, k)
  \ge \Adv{\sfe}_{\proto,\splitter,f,i}(\advA, \advS, k).
\]
\end{sloppypar}

% Even rounds
Now let $j\in[1..\ell]$ be even.
Adversary~$\advB_\even$ is constructed from~$\advA$ in a similar manner. (We
skip the details for brevity.) It emulates the same simulator~$\advS$ emulated
by~$\advB_\odd$. An identical argument yields
\[
  2 \cdot \Adv{\sfe}_{\proto_\even[f_{j}],\idsplitter,f_{j},i}(\advB_\even, \advS_\even, k)
  \ge \Adv{\sfe}_{\proto,\splitter,f,i}(\advA, \advS, k).
\]
Summing over all $j \in [1..\ell]$ yields the claim.

%\begin{comment}
\begin{figure}
  \oneCol{0.42}
  {
    \underline{$\advB_\odd(\pick, 1^k, f_j)$}\\[2pt]
      $(a, b, \sigma) \getsr \advA(\pick, 1^k, f)$\\
      $\veca \getsr \splitl(1^k, a)$;
      $\vecb \getsr \splitr(1^k, b)$;
      $\st, \vecy_0[0],\vecy_1[0] \gets \emptystr$\\
      \Foreach{h}{1}{j-1}\\
      \ind $u \gets \veca[h] \cat \vecy_0[h-1]$; $v \gets \vecb[h] \cat \vecy_1[h-1]$\\
      \ind if $h$ is odd then\\
      \ind\ind  $(\vecy_0[h], \vecy_1[h], \st) \gets f_h(u, v, \st)$\\
      \ind else \\
      \ind\ind $(\vecy_0[h], \vecy_1[h]) \gets f_h(u, v)$\\
      return $(\veca[j] \cat \vecy_0[j-1],
               \vecb[j] \cat \vecy_1[j-1], \st, (\sigma,j))$
    \\[6pt]
    \underline{$\advB_\odd(\guess, 1^k, \omega, (\sigma,j))$}\\[2pt]
      $d^\prime \getsr \bits$ \codecomment{``Guess'' our challenge bit.}\\
      $(L_\even^0, L_\even^1, L_\even^\O) \getsr \proto_\even.\init(1^k)$;
      $\st, y_0 \gets \emptystr$\\
      \Foreach{h}{1}{\ell}\\
      \ind $u \gets \veca[h] \cat \vecy_0[h-1]$; $v \gets \vecb[h] \cat  \vecy_1[h-1]$\\
      \ind if $d^\prime = 1$ then\\
      \ind\ind if~$h$ is odd then\\
      \ind\ind\ind $(\vecy_0[h], \vecy_1[h], \pi, \st) \getsr
          \proto_\odd[f_h](1^k, u, v, \st)$\\
      \ind\ind\ind $\omega_h \gets \View_{\proto_\odd[f_h],k}^i(\pi)$\\
      \ind\ind else \codecomment{$h$ is even}\\
      \ind\ind\ind $(\vecy_0[h], \vecy_1[h], \pi. \st^\prime) \getsr
          \proto_\even[f_h](1^k, u, v, \emptystr)$\\
      \ind\ind\ind $\omega_h \gets \View_{\proto_\even[f_h],k}^i(\pi)$\\
      \ind else \codecomment{$d^\prime = 0$}\\
      \ind\ind if~$h$ is odd then\\
      \ind\ind\ind $(\vecy_0[h], \vecy_1[h], \st) \gets f_h(u, v, \st)$\\
      \ind\ind\ind if $i = \Alice$ then
          $\omega_h \getsr \advS_\odd(1^k, u, |v|, \vecy_0[h], |\st|)$\\
      \ind\ind\ind else
          $\omega_h \getsr \advS_\odd(1^k, |u|, v, \vecy_1[h], |\st|)$\\
      \ind\ind else \codecomment{$h$ is even}\\
      \ind\ind\ind $(\vecy_0[h], \vecy_1[h]) \gets f_h(u,v)$\\
      \ind\ind\ind if $i = \Alice$ then
          $\omega_h \getsr \advS_\even(1^k, u, |v|, \vecy_0[h])$\\
      \ind\ind\ind else
          $\omega_h \getsr \advS_\even(1^k, |u|, v, \vecy_1[h])$\\
      $\omega_j \gets \omega$ \codecomment{Replace $j$-th round with our
      view.}\\
      $c^\prime \getsr \advA(\guess, \sigma, (\omega_1, \ldots,
        \omega_\ell))$\\
      return $c^\prime$
      \vspace{3pt}
  }
  \caption{Adversary~$\advB_\odd$ for proof of Theorem~\ref{thm:hyb}.
  The parties in the~$\proto_\odd$ protocol are given long-term inputs generated
  in $\advB_\odd$'s game.
  The parties in the~$\proto_\even$ protocol are given the long term inputs
  $(L_\even^0, L_\even^1, L_\even^\O)$ generated by~$\advB_\odd$.
  Note that we overload the syntax of the simulators by giving them strings
  where they expect vectors over strings.
  }
  \label{fig:pf-thm-hyb-adv}
\end{figure}
%\end{comment} % commenting this out to avoid extra pages appended at end

Due to space constraints, we refer the reader to our associated technical report
(anonymized for submission) for full proofs of Theorems 5.2 and 5.3.

%\begin{comment}
%%%%%%%%%%%%%%%%%%%%%%%%%%%%%%%%%%%%%%%%%%%%%%%%%%%%%%%%%%%%%%%%%%%%%%%%%%%%%%%
\subsection{Proof of Theorem~\ref{thm:gc}}\label{pf:thm-gc}
It suffices to prove the following:
Let~$\advA$ be an adversary and~$\advS_\garbler$ and~$\advS_\ot$ be simulators.
For each $i \in \{\Alice, \Bob\}$, there exist a pair of
adversaries~$\advB_\garbler$ and~$\advB_\ot$ and a simulator~$\advS$ such that
\[
  \Adv{\sfe}_{\proto,\idsplitter,f,i}(\advA, \advS, k) \le
    \Adv{priv}_{\garbler}(\advB_\garbler, \advS_\garbler, k) +
    \Adv{pfe}_{\proto_\ot}(\advB_\ot, \advS_\ot, k).
\]
Moreover, if~$\advA$, $\advS_\garbler$, and~$\advS_\ot$ are polynomial-time, then
so are~$\advB_\garbler$, $\advB_\ot$, and~$\advS$.

\begin{sloppypar} % to prevent equation from breaching column boundary
Our argument follows closely the proof of \cite[thm 14]{garbling}, except that
our goal is \sfe.
We begin with the case where $i=\Alice$. It suffices to specify~$\advB_\ot$ and
simulator~$\advS$ such that
\[
  \Adv{\sfe}_{\proto,\idsplitter,f,i}(\advA, \advS, k) \le
    \Adv{pfe}_{\proto_\ot}(\advB_\ot, \advS_\ot, k).
\]
% pick
During its $\pick$ phase, adversary~$\advB_\ot$ executes~$\advA$ on its $\pick$
phase, getting $(a,b, \sigma)$ in return.
It then samples random coins~$\coins$ and executes the garbling algorithm,
specifying~$\coins$ as the coins of execution. Let $(F, e, d)$ denote the
output.
Define~$g$ as the map $b \mapsto (X^{b_r}_{\ell+1}, \ldots, X^{b_r}_n)$ where
$\ell = |a|$, $r = |b|$, $n=\ell+r$, and $e = \tokens$.
Finally, adversary~$\advB_\ot$ returns $(g,b,\sigma)$.
% guess
Let~$\omega$ denote the (possibly-simulated) view given to~$\advB_\ot$ during
its $\guess$ phase. It constructs~$\Alice$'s view from~$\coins$, $\omega$, $Y =
\Ev(F, \Enl(e,a) \cat \Enr(e,b))$, and her initial state and executes~$\advA$
in its $\guess$ phase on the constructed view and its carry-over state~$\sigma$.
Finally, return whatever~$\advA$ returns.
In doing so, adversary~$\advB_\ot$ perfectly emulates~$\advA$ in its game
instantiated with a simulator~$\advS$ specified by the reduction and constructed
from~$\advS_\ot$.
\end{sloppypar}

We now turn to the case where $i=\Bob$.
% PFE adv
First, we construct the adversary~$\advB_\ot$ from~$\advS_\ot$ and~$\advA$ as
specified above.
% PRIV adv
We construct~$\advB_\garbler$ as follows.
During its $\pick$ phase, adversary~$\advB_\garbler$ executes~$\advA$ in its
$\pick$ phase on input $(1^k, f)$, getting $(a,b,\sigma)$ in return.
Adversary~$\advB_\garbler$ then returns $(f, (a,b), \sigma)$.
On input $(\guess, F, X, d, \sigma)$, adversary~$\advB_\garbler$ does as
follows. Since $\garbler$ is projective, there exists a sequence of tokens
$e=(Z_1^0,Z_1^1,\ldots,Z_n^0,Z_n^1)$ (unknown to~$\advB_\garbler$) such that $X
= A \cat B$ where $A = \Enl(e, a)$ and $B = \Enr(e, b)$.
Let $r = |b|$ and write~$B$ as the sub-sequence $(B_1, \ldots, B_r)$ of~$e$. For
each $i \in [1..r]$, let $X_i^{b_i} = B_i$ and let~$X_i^{1-b_i}$ be a random
token. Let $e^\prime = \tokens$.
Execute \[(\emptystr, B, \pi, \st^\prime) \getsr \proto_\ot(1^k, e^\prime, b,
\emptystr)\] and let $\omega^\prime \gets \View^{\Bob}_{\proto_\ot,k}(\pi)$.
Adversary~$\advB_\garbler$ constructs~$\Bob$'s view from~$F$, $A$,
$\omega^\prime$, and $\Bob$'s initial state, then executes~$\advA$ in its
$\guess$ stage with this view and its carry-over state~$\sigma$.
Finally, it outputs whatever $\advA$ outputs.

\begin{sloppypar} % to prevent equation from breaching column boundary
% Simulator
Lastly, we specify the simulator~$\advS$ as follows. On input $(1^k,
\len{\veca}, \vecb, \vecy)$ where $\veca[1] = a$, $\vecb[1] = b$, and $\vecy[1]
= f(a,b)$, the simulator executes $(F, X, d) \getsr \advS_\garbler(1^k, f,
f(a,b))$ and $\omega^\prime \getsr \advS_\ot(1^k, |g|, b, B)$, where~$g$ is the
map defined by $b \mapsto (X_{\ell+1}^{b_1}, \ldots, X_n^{b_r})$ for some set of
tokens $\tokens$ output by the garbling algorithm. (These lengths are known
since the simulator knows the function~$f$.). It constructs~$\Bob$'s view
from~$F$, $X$, $\omega^\prime$, and his initial state and returns it.
\end{sloppypar}

Since each of~$\advB_\ot$ and~$\advB_\garbler$ perfectly emulate~$\advA$ in its
game instantiated with~$\advS$, the claim holds.

%%%%%%%%%%%%%%%%%%%%%%%%%%%%%%%%%%%%%%%%%%%%%%%%%%%%%%%%%%%%%%%%%%%%%%%%%%%%%%%
\subsection{Proof of Theorem~\ref{thm:sgx}}\label{pf:thm-sgx}
The theorem statement in the body  is somewhat informal, since we formalize
\inddollar-security concretely instead of asymptotically. (In particular, we do
not associate a security parameter to the \inddollar~game.) Nevertheless, the
claim we prove here sufficient to establish the theorem for the natural
asymptotic definition.

We prove the following claim: Let~$\advA$ be an adversary.
For each $i \in \{\Alice, \Bob\}$, there exists an adversary~$\advB$ and a
simulator~$\advS$ such that for every~$k$, it holds that
\[
  \Adv{\wsfef}_{\proto,f,i}(\advA,\advS,k) \le \Adv{\inddollar}_\symenc(\advB)
\]
Moreover, adversary~$\advB$ has about the same runtime as~$\advA$ and~$\advS$
runs in constant time.

We start again with $i=\Bob$. $\Bob$'s view consists only of its initial state
and its coins for encrypting its message to~$\Alice$. Clearly there exists a
simulator~$\advS$ such that $\Adv{\wsfef}_{\proto,f,i}(\advA,\advS,k) = 0$.

Now consider $\Alice$'s view. Fix a positive integer~$k$.
Adversary~$\advB$ first executes $(a, b, \st, \sigma) \getsr \advA(\pick,
1^k, f)$ and asks~$b$ of its oracle, getting~$x$ in response. It then constructs
$\Alice$'s view~$\omega$ from her initial state, $x$, and~$y$ where $(y, \st^\prime) =
f(a,b,\st)$. It then executes $c^\prime \getsr \advA(\guess, \omega, \sigma)$
and returns $c^\prime$.

If~$\advB$ has an encryption oracle, then it perfectly emulates~$\advA$ in its
game with~$c=1$ as the challenge bit. If~$\advB$ is a~$\$(\cdot)$ oracle, then
it perfectly emulates~$\advA$ in its game with~$c=0$ and a
simulator~$\advS$ specified by the reduction. Hence,
\[
  \Prob{K \getsr \kg: \advB^{\enc_K(\cdot)} \outputs \true} =
  \Prob{\Exp{\wsfef}_{\proto,f,i}(\advA,\advS,k) \outputs \true | c = 1}
\]
and
\[
  \Prob{\advB^{\$(\cdot)} \outputs \false} =
  \Prob{\Exp{\wsfef}_{\proto,f,i}(\advA,\advS,k) \outputs \true | c = 0}
\]
where $i=\Alice$. Finally,
\begin{equation*}
  \begin{aligned}
    \Adv{\inddollar}_\symenc(\advB) &=
      2\cdot \Prob{\Exp{\wsfef}_{\proto,f,i}(\advA,\advS,k) \outputs \true} - 1\\
      &= \Adv{\wsfef}_{\proto,f,i}(\advA,\advS,k) \,.
  \end{aligned}
\end{equation*}
%\end{comment}
\end{techreport}

%\end{techreport}

\end{document}